\documentclass[article, 10pt]{IEEEtran}

\usepackage{amsmath}
\usepackage{amsfonts}
\usepackage{amssymb}
\usepackage{graphics}
\usepackage{empheq}
\usepackage{pstricks}
\usepackage{array}
\usepackage{float}
\usepackage{epsf}
\usepackage{graphicx}
\usepackage{subfigure}
\usepackage{multirow}
\usepackage{cite}
\usepackage{physics}
\usepackage{acronym}
\usepackage{comment}
\usepackage{upgreek}
\usepackage{subcaption}
\usepackage{nicefrac} 
\usepackage[font=small]{caption}
\usepackage{comment}
\usepackage{pifont}

\newcommand{\wo}{w_o}

\newcommand{\Dfi}{{\rm \Delta} f_{i}}
\newcommand{\rocofH}{{\rm RoCoF}|_{t=0_+}}
\newcommand{\rocofi}{{\rm RoCoF}_{i}}
\newcommand{\rocofavg}{\overline{{\rm RoCoF}_{i}}}
\newcommand{\rocofmax}{{\rm RoCoF}_{{\max},i}}
\newcommand{\rocofra}{{\rm RoCoF}_{\rm r1}}
\newcommand{\rocofrb}{{\rm RoCoF}_{\rm r2}}
\newcommand{\sigmaf}{\sigma_f}
\newcommand{\Dsigmaf}{{\rm \Delta} \sigmaf}

\acrodef{rocof}[RoCoF]{Rate of Change of Frequency}

\acrodef{aips}[AIPS]{All-Island Power System}
\acrodef{aus}[AUS]{Australia}
\acrodef{tas}[TAS]{Tasmania}
\acrodef{gb}[GB]{Great Britain} 
\acrodef{agc}[AGC]{Automatic Generation Control}
\acrodef{pfc}[PFC]{Primary Frequency Control}
\acrodef{aemo}[AEMO]{Australian Energy Market Operator} 
\acrodef{der}[DER]{Distributed Energy Resource}
\acrodef{ibr}[IBR]{Inverter-Based Resource}
\acrodef{pv}[PV]{Photovoltaic}
\acrodef{tso}[TSO]{Transmission System Operator}
\acrodef{ce}[CE]{Continental Europe}
\acrodef{afrr}[aFRR]{Automatic Frequency Restoration Reserve}
\acrodef{fcr}[FCR]{Frequency Containment Reserve}
\acrodef{ffr}[FFR]{Fast Frequency Response}
\acrodef{entsoe}[ENTSO-E]{European Network of TSOs for Electricity}
\acrodef{dm}[DM]{Dynamic Moderation}
\acrodef{dr}[DR]{Dynamic Regulation}
\acrodef{neso}[NESO]{National Energy System Operator}
\acrodef{res}[RES]{Renewable Energy Sources}
\acrodef{nem}[NEM]{National Electricity Market}
\acrodef{fat}[FAT]{Full Activation Time} 
\acrodef{bess}[BESS]{Battery Energy Storage Systems}
\acrodef{sir}[SIR]{Synchronous Inertial Response}
\acrodef{lsi}[LSI]{Largest Single Infeed}
\acrodef{lso}[LSO]{Largest Single Outfeed}
\acrodef{fpd}[FPD]{Frequency Probability Distribution}
\acrodef{ic}[IC]{Interconnector}
\acrodef{por}[POR]{Primary Operating Reserve}
\acrodef{sor}[SOR]{Secondary Operating Reserve}
\acrodef{tor1}[TOR1]{Tertiary Operating Reserve 1}
\acrodef{tor2}[TOR2]{Tertiary Operating Reserve 2}
\acrodef{mfrr}[mFRR]{Manual Frequency Restoration Reserve}
\acrodef{rr}[RR]{Replacement Reserve}
\acrodef{rrd}[RRD]{Replacement Reserve Desynchronized}
\acrodef{rrs}[RRS]{Replacement Reserve Synchronized}
\acrodef{of}[OF]{Over-frequency}
\acrodef{uf}[UF]{Under-frequency}
\acrodef{dc}[DC]{Dynamic Containment}
\acrodef{hvdc}[HVDC]{High Voltage Direct Current}
\acrodef{ercot}[ERCOT]{Electric Reliability Council of Texas}
\acrodef{cps}[CPS]{Control Performance Standard}
\acrodef{pfr}[PFR]{Primary Frequency Regulation}
\acrodef{dso}[DSO]{Distribution System Operator}
\acrodef{frcr}[FRCR]{Frequency Risk and Control Report}
\acrodef{sfc}[SFC]{Secondary Frequency Control}
\acrodef{sfr}[SFR]{Secondary Frequency Regulation}

\begin{document}

\title{A Comprehensive Approach to Evaluate Frequency Control Strength of Power Systems}

\author{Taulant K{\"e}r{\c c}i,~\IEEEmembership{Senior Member,~IEEE},
  and Federico~Milano,~\IEEEmembership{Fellow,~IEEE}
  \thanks{F.~Milano is with School of Electrical and Electronic
    Engineering, University College Dublin, Belfield Campus, Dublin 4,
    Ireland.  e-mail: federico.milano@ucd.ie.}
  \thanks{This work was partially supported by Research Ireland by
    funding Taulant K\"er\c{c}i under NexSys project, Grant
    No.~21/SPP/3756; and by the Sustainable Energy Authority of
    Ireland (SEAI) by funding Federico Milano through FRESLIPS
    project, Grant No.~RDD/00681.}
}

\maketitle

\begin{abstract}
  This paper introduces the concept of ``frequency control strength'' as a novel approach to understand how different real-world power systems compare to each other in terms of effectiveness and performance of system-wide frequency control.  It presents a comprehensive comparison, based on measurement data, of the frequency control strength of four real-world, renewable-based, synchronous islands power systems, namely Great Britain (GB), All-Island power system (AIPS) of Ireland, and Australia (AUS) mainland and Tasmania (TAS).  The strength is evaluated by means of different frequency quality metrics.  The common understanding is that the bigger the capacity of a power system, the bigger its robustness with respect to events and contingencies.  Here we show that this is not always the case in the context of frequency control.  In fact, our study shows that mainland AUS shows the highest frequency control strength during normal operating conditions, whereas the AIPS shows the highest relative frequency control strength for abnormal system conditions.  The strength is, in particular, greatly influenced by different regulatory requirements and different system/ancillary services arrangements in each jurisdiction.  The paper also provides possible mitigations to improve frequency control strength through grid codes and market rules.
\end{abstract}


\section{Overview}
\label{sec:05:intro}

Frequency control of power systems is an emerging area of research due to the increasing penetration of variable \ac{res} such as wind and solar \ac{pv} generation \cite{kerci2024emerging, BEVRANI2021107114}.   For example, as the inertia decreases with the displacement of conventional synchronous generators there is a concern that frequency excursions become faster, and therefore the likelihood of instability occurring earlier increases  \cite{9286772}.  While there are significant ongoing efforts from both academia and industry on how to best deal with the frequency control challenges (during both normal and abnormal system conditions), it is not clear how different real-world \ac{res}-dominated power systems compare in terms of the ``strength'' of frequency control.  This paper fills this gap by comparing the strength of frequency control of four real-world synchronous islands power systems namely the \ac{gb}, \ac{aips}, and \ac{aus} mainland,\footnote{In the reminder of this work \ac{aus} indicates the combined transmission system of Queensland, New South Wales, Victoria, and South Australia \cite{aemo3}.} and \ac{tas}.  The frequency control strength of each system is evaluated by means of various metrics of frequency quality.    

It is generally understood that small (islanded) synchronous power systems exhibit rapid and substantial frequency excursions compared to large power systems following typical generation loss \cite{10015191, overview}.  For instance, when comparing the different frequency control of \ac{gb}, \ac{aips}, and \ac{aus} power system, reference \cite{nedd2020operating} states: \textit{Ireland is chosen because its power system has, arguably, greater challenges than \ac{gb}'s, owing to its small size, limited interconnection and high penetration of wind.  \ac{aus} was chosen because of its National Electricity Market's comparable size and renewable penetration to the \ac{gb} grid.} The measurement data discussed in this paper, however, demonstrate that size may not necessarily mean greater or lower frequency control challenges.  Specifically, despite \ac{gb} being bigger/much bigger than \ac{aips}, \ac{aus}, and \ac{tas} systems (see next section), it appears, based on this study, to face greater frequency regulation challenges.  We elaborate on these differences by means of the concept of \textit{frequency control strength}.

There is currently no commonly accepted definition of ``frequency strength'' and how it may relate and overlap with system strength definition(s) \cite{badrzadeh2024power}.  For instance, reference \cite{en16041842} suggests that frequency strength exhibits its meaning in two dimensions, namely the inertia support capability, which defines the initial \ac{rocof} (see equation \eqref{eq:05:rocof2} below) after an active power disturbance, and the \ac{pfc} capability, which defines the amount of active power that the power system can absorb or release during the frequency deviation.  However, this definition is incomplete as it assumes that frequency strength only deals with contingency events and does not include the ability of power systems to counteract frequency deviations during normal operating conditions.  

Motivated by this confusion and gap in the literature we propose the following definition of frequency strength:

\vspace{4mm}

\fbox{%
\begin{minipage}{0.9\linewidth}
\parbox{\textwidth}{%
  \textbf{Frequency strength} is the ability of power systems to resist and control changes in frequency during normal and abnormal operating conditions.%
  }
\end{minipage}%
}

\vspace{4mm}

In this context, references \cite{9712369, 10252953} propose real-time frequency strength evaluation indices based on a unified transfer function\index{transfer function} structure to theoretically quantify the frequency strength in terms of nadir and the average \ac{rocof}.  In the same vein, a frequency security index to evaluate power system frequency performance is proposed in \cite{9309179}.  These metrics, however, aim at quantifying frequency strength only during abnormal operating conditions.  On the other hand, reference \cite{10210321} focuses only on normal system conditions and proposes frequency regulation performance requirement constraints into a generation planning model to ensure frequency quality.  Frequency quality is also studied in \cite{MOHAMMADI2024101359} with a focus on sub-15-min and sub-1-hr time scales and the highest data resolution used is 10 s.

References \cite{HONG20211057, 10273444} provide an overview of frequency control challenges in the \ac{gb} and \ac{aus} power systems, respectively.  Reference \cite{HOMAN202156} discusses the efficacy of the proposed new frequency response services in \ac{gb} using a month-long case study.  Similarly, references \cite{HOMAN202063, HOMAN2021116723} study the volatility of \ac{gb} frequency using 1 s frequency resolution data and demonstrate the existence of the relationship between increased frequency events and \ac{res} penetration, and rate of change of demand, respectively.  Reference  \cite{wen2023non} utilizes real-world frequency data of different countries in Asia, \ac{aus}, and Europe and compares the statistical properties of the frequency, that is, asymmetry of \ac{fpd}.  

This paper compares the strength of frequency control under normal and abnormal operating conditions of four power systems that are at the forefront of the integration of \ac{res}, namely, \ac{gb}, \ac{aips}, \ac{aus} and \ac{tas}.  
Novel contributions of this work are as follows.
\begin{itemize}
\item A comprehensive review of the main characteristics of \ac{gb}, \ac{aips}, \ac{aus} and \ac{tas} systems and an overview of their frequency control.
\item Novel metrics to evaluate and enable a systematic comparison of frequency control strength on actual historical frequency measurements/events.
\item It highlights that, overall, the \ac{aus} power system shows higher frequency regulation strength, while the \ac{aips} power system shows the ``strongest'' frequency performance when it comes to arresting frequency and \ac{rocof} relative to its size.
\item A set of possible mitigations to improve frequency performance (i.e., grid codes and market rules).
\end{itemize}

\section{System Characteristics and Frequency Control}
\label{sec:05:comp}

This section provides a detailed comparison, from the frequency control point of view, of the main characteristics of the \ac{gb}, \ac{aips}, \ac{aus}, and \ac{tas} power systems.  These systems are at the forefront internationally to integrate pioneering levels of \ac{res}.  Table \ref{tab05:param1} compiles all the relevant information.  Note the different sizes of the systems in terms of peak demand and inertia floors.\index{inertia floor}  Specifically, the \ac{gb} system is, from a peak demand and inertia perspective, between 5.2-6.3, 1.3-3.3, and 22-37 times bigger than \ac{aips}, \ac{aus} and \ac{tas} systems, respectively.

On the other hand, the four power systems are very similar in terms of instantaneous \ac{res} penetration and electricity met by \ac{res}.  Concerning the \ac{pfc} provision, the \ac{gb} system differs from \ac{aips}, \ac{aus}, and \ac{tas} systems as it procures/pays units to reserve headroom (but \ac{pfc} bids are mandatory) \cite{nedd2020operating}.  Motivated by a decline in the frequency control performance under normal conditions, in 2020 the Australian Energy Market Commission made a final (mandatory) rule to require all generators (scheduled and semi-scheduled) in the \ac{nem} to provide \ac{pfc} response (droop-based) with narrow dead-band ($\pm15$ mHz) and thus support the secure operation of the power system.  This change led to a significant improvement in frequency quality in the \ac{nem} which will be discussed in detail in the case study section.  Next, all four systems have similar requirements for  dead-band ($\pm15$ mHz) and droop (3-5\%).  Regarding other relevant services, \ac{aips} includes an \ac{sir} service while the rest do not.  This is not the case for the \ac{ffr} service where all four systems have recently introduced it with a \ac{fat} between 0.15-2 s.  Note that \ac{ffr} is a class of \ac{pfc} that is more needed at low levels of synchronous inertia \cite{milano_2018}.  \ac{bess} are one of the most favorable candidates to provide the \ac{ffr} service due to their fast responsive time and flexibility of operation \cite{akram2020review}.  With respect to \ac{agc}, both \ac{gb} and \ac{aips} do not currently implement it while \ac{aus} and \ac{tas} do.

\begin{table*}[t!]
  \centering
  \caption[Main characteristics of four power systems]{Main characteristics of \ac{gb}, \ac{aips}, \ac{aus} and \ac{tas} power systems.}
  \label{tab05:param1}
  \resizebox{1.0\linewidth}{!}{
  \begin{tabular}{ccccc}
    \hline
    Item & \ac{gb} & \ac{aips} & \ac{aus} & \ac{tas} \\
    \hline
    Peak demand [GW] & 44 & 7.5 & 34 & 2  \\
    Inertia floor\index{inertia floor} [GWs] & 120 & 23 & 35.8 & 3.2  \\
    Instantaneous \ac{res} [\%] & 87.6 & 75 & 72.1 & 100 \\
    Electricity from \ac{res} [\%] & 51 & 42 & 39.4 & 93.4 \\
    \ac{pfc} provision & Market/Mandatory & Mandatory & Mandatory & Mandatory \\
    dead-band [mHz] & $\leq \pm 15$ & $\leq \pm 15$ & $\leq \pm 15$ & $\leq \pm 15$ \\
    Droop [\%] & 3-5 & 3-5 & $\leq$ 5 & $\leq$ 5 \\
    \ac{sir} & No & Yes & No & No  \\
    \ac{ffr} & Yes (1 s) & Yes (0.15-2 s) & Yes (0.5-1 s) & Yes (0.5-1 s) \\
    \ac{agc} & No & No & Yes & Yes \\
    \ac{rocof} ride-through requirement [Hz/s] & 0.125 - 1 & 1 & 1 & 3 \\
    \ac{lsi} loss [MW] & 1400 & 500 & - & 144 \\
    \ac{lso} loss [MW] & 1400 & 500 & - & 144 \\
    Number of \acp{ic} (ac/\ac{hvdc}) & 10 & 3 & 6 & 1 \\
    \ac{ic} (ac/\ac{hvdc}) ramp rate [MW/min] & 100 & 5 & 40 & 40 \\
    Dispatch model & Self & Central & Central & Central \\
    Flexibility markets & Mature & Early stage & Early stage & Early stage \\
    \hline
  \end{tabular}}
\end{table*}

With regard to \ac{rocof} ride-through requirement, all systems have a limit ranging from 0.125 Hz/s in \ac{gb} (the most sensitive \ac{rocof} protection on the \ac{gb} system) to 3 Hz/s in \ac{tas}.  This makes sense since \ac{tas} is a much smaller system and thus needs to deal with higher \acp{rocof}.  In terms of \ac{lsi} and \ac{lso} loss, 
\ac{neso} operates and designs the system assuming a 1400 MW loss (the capacity of an \ac{ic}) to ensure the resulting \ac{rocof} would not exceed 0.5 Hz/s while \ac{aips} a 500 MW and \ac{tas} a 144 MW loss (\ac{aus} does not include this criterion).  There is also a significant difference in the number of ac and dc \acp{ic}.  Specifically, this number ranges from 10 for \ac{gb} to just 1 for \ac{tas}.  Note that while more \acp{ic} between countries mean better market integration, it could also have the negative effect on frequency control if several \acp{ic} ramp at the same time.  In this context, \ac{gb} has a much higher \ac{ic} ramp rate (100 MW/min) compared to \ac{aips} (5 MW/min), and \ac{aus}/\ac{tas} (40 MW/min).  For comparison, the \ac{aips} system expects to have a combined \ac{ic} ramp rate of 40 MW/min by 2030 which has to be compared to the current limit of 15 MW/min.  These aspects are critical to frequency variations and are discussed below.

Table \ref{tab:05:control} summarizes all the relevant frequency control products of the selected jurisdictions.  More specifically, Table \ref{tab:05:control} suggests that reserve products in \ac{aips} namely \ac{por}, \ac{sor}, \ac{tor1}, \ac{tor2}, and \ac{rr} are designed and procured to deal with \ac{uf} events (i.e, upward reserves), while in \ac{gb} and \ac{aus}/\ac{tas} reserves are procured to deal with both under and \ac{of} events (\ac{pfc}-based upward/downward reserves).  Note that \ac{dc}/\ac{dr}/\ac{dm} services (all with a $\pm15$ mHz dead-band and predominately being provided by \ac{bess}) are faster than the two legacy response products mandatory frequency response and static firm frequency response services.  

\begin{table*}[htb]
  \centering
  \caption[Frequency services of four power systems]{Comparison of frequency services of \ac{gb}, \ac{aips} and \ac{aus}/\ac{tas} systems.} 
  \label{tab:05:control}
\begin{tabular}{cccccc}
\hline
Jurisdiction & Service & Direction & FAT & Purpose \\
& & (Upward/Downward)& [s] &    \\
\hline
\ac{aips} & \ac{ffr} & Upward & 0.15-2 & Post-fault contingency\\
\ac{aips} & POR & Upward & 5 & Post-fault contingency\\
\ac{aips} & SOR & Upward & 15 & Post-fault contingency\\
\ac{aips} & TOR1 & Upward & 90 & Post-fault contingency\\
\ac{aips} & TOR2 & Upward & 300 & Post-fault contingency\\
\ac{aips} & RR & Upward & 1200 & Post-fault contingency\\
\ac{gb} & \ac{dc} & Upward/Downward & 1  & Post-fault contingency \\
\ac{gb} & \ac{dm} & Upward/Downward & 1 & Pre-fault continuous \\
\ac{gb} & \ac{dr} & Upward/Downward & 10 &  Pre-fault continuous \\
\ac{gb} & Mandatory frequency response & Upward/Downward & $<$ \ac{dc}/\ac{dm}/\ac{dr} &  Post-fault continuous \\
\ac{gb} & Static firm frequency response & Upward & 10-30 &  Post-fault contingency \\
\ac{gb} & Commercial frequency response & Varies & Varies &  Post-fault contingency \\
\ac{gb} & Slow reserve & Upward/Downward & 900 &  Post-fault contingency \\
\ac{gb} & Quick reserve & Upward/Downward & 60 &  Pre-fault continuous \\
\ac{gb} & Fast reserve & Upward/Downward & 120 &  Pre-fault continuous \\
\ac{gb} & Balancing reserve & Upward/Downward & 600 &  Pre-fault continuous \\
\ac{aus}/\ac{tas} & \ac{ffr} & Upward/Downward & 0.1-1  & Post-fault contingency\\
\ac{aus}/\ac{tas} & Fast reserve & Upward/Downward & 6  & Post-fault contingency\\
\ac{aus}/\ac{tas} & Slow reserve & Upward/Downward & 60  & Post-fault contingency\\
\ac{aus}/\ac{tas} & Delayed reserve & Upward/Downward & 300  & Post-fault contingency\\
\ac{aus}/\ac{tas} & \ac{agc} & Upward/Downward & 300  & Pre-fault continuous\\
    \hline
  \end{tabular}
\end{table*}

One may argue, and we certainly agree, that there is no need for so many frequency reserve products.  For instance, \ac{entsoe} has defined and recommended the use of four standard reserve products namely \ac{fcr}, \ac{afrr}, \ac{mfrr}, and \ac{rr} \cite{sogl}.  In addition to these products, \acp{tso} operating low-inertia grids might need to introduce an \ac{ffr} product.

\section{Frequency Control Strength Metrics}
\label{sec:05:metrics}

\subsection{Nadir/Zenith and Minutes-based Metrics}
\label{sec:05:quality}

Minutes-based metrics such as ``minutes outside the normal operating band'' are, more often than not, used by \ac{entsoe} as a measure of long-term/annual frequency quality.  Table \ref{tab05:param} presents various frequency quality parameters for the four power systems.  \acp{tso} continuously monitor and periodically report on these indices \cite{aemo3}.  In particular, it is worth pointing out that the standard frequency range in \ac{gb} and \ac{aips} is $\pm 200$ mHz while in \ac{aus} and \ac{tas} is $\pm150$ mHz.  In addition, \ac{aips} and \ac{gb} have a target to maintain frequency within an even tighter range namely $\pm 100$ mHz for $\geq$ 98\% of the time, while \ac{aus} and \ac{tas} have a target to keep frequency within $\pm 150$ mHz for $\geq 99\%$ of the time.  We apply these metrics to operational data to evaluate and quantify the strength of frequency control under both normal and abnormal system conditions.

\begin{table*}[t!]
  \centering
  \caption[Frequency quality parameters of four power systems]{Relevant frequency quality parameters of the \ac{ce}, \ac{gb}, \ac{aips}, \ac{aus} and \ac{tas} power systems.}
  \label{tab05:param}
  \resizebox{1.0\linewidth}{!}{
  \begin{tabular}{cccccc}
    \hline
    \multirow{2}{*}{Parameter}  & \multirow{2}{*}{\ac{ce}}  & \multirow{2}{*}{\ac{gb}} & \multirow{2}{*}{\ac{aips}} & \ac{aus} & \multirow{2}{*}{\ac{tas}}  \\
    & & & & (mainland) & \\
    \hline
    Standard frequency range [mHz]  & $\pm50$  & $\pm200$  & $\pm200$ & $\pm150$ & $\pm150$ \\ 
    Frequency key performance indicator [mHz]  & -  & $\pm100$ ($\ge 98\%$)  & $\pm100$ ($\ge 98\%$)  & $\pm150$ ($\ge99\%$) & $\pm150$ ($\ge 99\%$) \\ 
    Maximum instantaneous frequency deviation [mHz] & $800$  & $800$  & $1000$ & $1000$ & $2000$ \\
    Maximum steady-state frequency deviation [mHz] & $200$  & $500$ & $500$ & $500$ & $500/1000$ \\
    Time to restore frequency [min] & $15$ & $15$  & $15$ & - & -\\
    Frequency restoration range [mHz] & not used & $\pm200$ & $\pm200$ & - & -\\
    Maximum number of minutes & \multirow{2}{*}{15,000} & \multirow{2}{*}{15,000} & \multirow{2}{*}{15,000} & \multirow{2}{*}{-} & \multirow{2}{*}{-} \\
    outside the standard frequency range \\
    Minutes outside normal operating frequency band& \multirow{2}{*}{-} & \multirow{2}{*}{-} & \multirow{2}{*}{-} & \multirow{2}{*}{$\leq$ 5 (5)} & \multirow{2}{*}{$\leq 5$ (10)} \\
    during normal (abnormal) system conditions \\
    \hline
  \end{tabular}}
\end{table*}

Similarly, \acp{tso} operate and design power systems based on predefined maximum instantaneous frequency deviations namely nadir and zenith, and if these limits are exceeded then defense measures may be in place (for example, load/generation shedding) to avoid system blackouts.

\subsection[Frequency Deviation and RoCoF-based Metrics]{${\rm \Delta} f$ and \ac{rocof}-based Metrics}
\label{sec:05:rocofbased}

Frequency stability is generally evaluated based on three key metrics namely \ac{rocof}, nadir and zenith \cite{hurtado2024stability}.  \ac{rocof} measures how fast frequency changes following imbalances between generation and demand.  \ac{rocof} is important during frequency transients as it is used by protections such as, for example, loss-of-mains protection settings by generators.  

The initial \ac{rocof} is calculated as follows:
\begin{equation}
  \label{eq:05:rocof2}
  \rocofH = \frac{{\rm \Delta} p_{\rm imbalance}}{p_{\rm load}} \, \frac{\wo}{4 \pi H_{\rm agg}} \, ,
\end{equation}
where ${\rm \Delta} p_{\rm imbalance}$ is the size of the infeed/outfeed outage event, $p_{\rm load}$ is the current system load consumption; $H_{\rm agg}$ is the system aggregated inertia constant; $\omega_o$ is the synchronous reference frequency.
$\rocofH$ is used by the four \acp{tso} that we examine in this paper to determine the minimum inertia floors\index{inertia floor} to maintain \ac{rocof} within limits \cite{NESOrocof, 10253224, 10273444}.  

This way to calculate the \ac{rocof} is motivated by the swing equation of synchronous machines but gives only a snapshot and only at the initial instant after a major event.  Moreover, $\rocofH$ is a conservative calculation as it neglects the \ac{ffr} being provided in the inertial time frame.  Thus, it appears useful to have additional information over a mobile window in the seconds after the event.  With this aim, we define first the frequency window of interest as:
\begin{align}
  \label{eq:05:deltaf} 
  \Dfi(t) &= \big | f(t) - f(t-i) \big | \, , 
\end{align}
where $f(t)$ is a suitable estimation of the instantaneous frequency at time $t$.
As it stands, $\Dfi$ is a useful metric for normal operating conditions and we will use it in the remainder of this paper to compare long-term frequency deviations of real-world power systems. 

We are now ready to define \ac{rocof}-based metrics that are complementary to \eqref{eq:05:rocof2}, as follows: 
\begin{align}
  \label{eq:05:rocof1} \rocofi(t) &= \frac{\Dfi(t)}{i} \, , \\
  \label{eq:05:averagerocof} \rocofavg(t) &= \frac{1}{N} \sum \limits_{h=0}^{N-1} \rocofi(t-h{\rm \Delta} t) \, , \\
  \label{eq:05:rocofmax} \rocofmax(t) &= \underset{h = 0, \dots, N-1}{\max} \big \{ \rocofi(t - h {\rm \Delta} t) \big \} \, .
\end{align}

Equation \eqref{eq:05:rocof1} is \ac{rocof} calculated for a given period $i$, whereas equations \eqref{eq:05:averagerocof} and \eqref{eq:05:rocofmax} are the average and maximum \acp{rocof}, respectively, calculated over the time period $i$ with ${\rm \Delta} t = i/N$.  While these \ac{rocof}-based metrics can be applied during both normal and abnormal system conditions, these three metrics best characterize abnormal operating conditions.

The metrics above are ``absolute.''  To be able to compare the strength of frequency control of different power systems it is also useful to consider ``relative'' \ac{rocof}-based calculations that can take into account the size of the power systems.  With this aim, we propose two relative metrics, as follows:
\begin{align}
  \label{eq:05:deltaw} \rocofra &= \rocofmax \, \frac{p_{\rm total}}{{\rm \Delta} p_{\rm imbalance}} \, ,\\
  \label{eq:05:deltaw1} \rocofrb &= \rocofmax \, \frac{
  p_{\rm conv}}{{\rm \Delta} p_{\rm imbalance}} \, ,
\end{align}
where $p_{\rm total}$ is the total active power in the system; and $p_{\rm conv}$ is the amount of conventional generation in the system.  These metrics can be interpreted as follows.  Two systems should have same $\rocofra$ ($\rocofrb$) if their inertia and control are equally proportional to $p_{\rm total}$ ($p_{\rm conv}$).  For example, the product $\rocofmax \cdot p_{\rm total}$ should be the same for two systems controlled in the same way.  On the other hand, a weak system will show a higher $\rocofra$ than a strong one.

\subsection{Frequency Standard Deviation-based Metrics}
\label{sec:05:std}

Frequency standard deviation-based metrics are widely and for a long-time used in power systems.  For instance, the \acp{tso} in the US utilize the standard deviation of the frequency, say, $\sigmaf$ (calculated based on 1-minute frequency deviation averages over a year), as a long-term frequency \ac{cps}1 \cite{7084686}.  In the case of \ac{ercot},\index{ERCOT} \ac{cps}1 compliance is assumed if $\sigmaf$(year) $\leq$ 30 mHz \cite{7084686}.

In this context, we recently proposed a new metric based on $\sigmaf$ to calculate and measure the asymmetry of the frequency distribution ($\Dsigmaf$) in power systems \cite{kerci2024asymmetry}.  

In an ideal scenario $\Dsigmaf$=0.  However, this is impossible in practice due to losses and nonlinearity.  It makes sense thus to use the asymmetry index as a measure of strength of frequency control as a perfect frequency control would lead to $\Dsigmaf=0$.

\section{Real-world System Data}
\label{sec:05:case}

In this section, we apply the metrics described in the previous section to real-world data of the four power systems for both normal and abnormal (contingency) operating conditions.  These have been made publicly available by the relevant \acp{tso}.  In particular, and if not otherwise stated, we use 1 s frequency resolution time series data for \ac{gb} and \ac{aips} and 4 s for \ac{aus} and \ac{tas} power systems.  For this reason, and where possible, we calculate the relevant frequency control strength metrics at 4 s resolution for \ac{gb} and \ac{aips} to allow a direct comparison with \ac{aus} and \ac{tas} systems.

\subsection{Normal System Conditions}
\label{sec:05:normal}

This section focuses on normal operating conditions and applies various minutes-based, $\Dfi$-based and $\sigmaf$-based metrics to quantify and evaluate frequency strength of the \ac{gb}, \ac{aips}, \ac{aus} and \ac{tas} power systems.

\subsubsection{Minutes-based comparison}
\label{sec:05:qualityresults}

The first comparison that we look at are the minutes outside the relevant standard frequency ranges.  For this, we select year 2023 and present all the results in Table \ref{tab05:param}.  For illustration and comparison purposes, we present these minutes for the \ac{ce} power system as well which has a peak demand of around 440 GW and thus approximately ten times bigger than \ac{gb}.  While the standard frequency range for \ac{gb} and \ac{aips} is $\pm 200$ mHz, we calculate the minutes outside $\pm100$ mHz as well as this is an even tighter frequency band that the \acp{tso} of \ac{gb} and \ac{aips} systems aim at maintaining and reporting on a continuous basis.  On the other hand, while the standard frequency range for \ac{aus} and \ac{tas} is $\pm 150$ mHz, we calculate the minutes outside $\pm100$ mHz as well to allow a direct comparison with \ac{gb} and \ac{aips} power systems.  Regarding the \ac{ce} statistics, we show the minutes outside $\pm 50$ mHz. 

Table \ref{tab:05:quality} suggests that despite \ac{ce} being the biggest power system, it is the one that has exceeded, for the first time, its annual frequency quality target, namely, frequency outside the $\pm 50$ mHz range for more than 15,000 minutes.  Table \ref{tab:05:quality} also suggests that, by far, frequency in \ac{gb} is spending much more minutes outside the $\pm 200$ mHz (710.6) and $\pm 100$ mHz (80,131.36) ranges compared to the \ac{aips}, \ac{aus} and \ac{tas} power systems.  It should be noted, however, that \ac{gb} is still within frequency quality limits in terms of minutes outside standard frequency range ($\pm 200$ mHz) in Table \ref{tab05:param} (15,000 minutes).  As a matter of fact, in 2014 frequency was inside the $\pm 100$ mHz range for 94\% of the time (or approximately 534 hours outside) compared with 90\% in 2021 (or approximately 832 hours outside) \cite{NGESO}.  It seems, though, that frequency spent even more hours outside limits during 2023 (approximately 1335) compared to 2014 (534) and 2021 (832).  This is interesting considering that \ac{gb} recently launched two new \ac{pfr} products namely \ac{dr} and \ac{dm} to tackle the challenge of frequency regulation.  It appears, though, that frequency regulation is, as expected by \ac{neso}, still a major challenge to be addressed \cite{NGESO}.  In fact, \ac{neso} anticipates \textit{``this exposure to increase in the future as the system is getting more volatile (more renewable connected, low inertia, large uncertainty)}, and suggests \textit{the need for faster response and reserve products''} \cite{NGESO}.

\begin{table*}[t!]
  \centering
  \caption[Frequency quality in 2023 for four power systems]{Frequency quality in 2023 for the power systems of \ac{ce}, \ac{gb}, \ac{aips}, \ac{aus} and \ac{tas}.}
  \label{tab:05:quality}
  \begin{tabular}{lccccc}
    \hline
    Minutes outside & \ac{ce} & \ac{gb} & \ac{aips} & \ac{aus} & \ac{tas}    \\
    \hline
    $\pm 50$ mHz & 15,389 & - & - & -  & - \\
    $\pm 200$ ($\pm 100$) mHz & - & 710.6 (80,131.36) & (3.56) 6,796 & - & - \\
    $\pm 150$ ($\pm 100$) mHz & - & - & - & 3.4 (134.26) & 2,074.6 (7,157.66) \\
    \hline
  \end{tabular}
\end{table*}

Looking at the \ac{aips}, \ac{aus} and \ac{tas} statistics in Table \ref{tab:05:quality}, one can see that \ac{aus} outperforms the rest of power systems (only 3.4 minutes outside $\pm 150$ mHz and 134.26 minutes outside $\pm 100$ mHz in 2023).  This could be explained by the much bigger size of the \ac{aus} system compared to the \ac{aips} and \ac{tas} systems and the fact that the \ac{aus} system utilizes an \ac{agc} compared to the \ac{aips} system.  What is interesting, though, is that despite being slightly a smaller system than \ac{gb}, the \ac{aus} system shows a significant better frequency performance during normal operating conditions.  One of the main differences with \ac{gb} relate to the fact that the \ac{pfc} provision ($\pm 15$ mHz dead-band and droop/proportional response) is fully mandatory in the \ac{aus} system whereas that is not the case in \ac{gb} (see Table \ref{tab05:param1}).  This can be seen as the frequency regulation task is distributed among many units in \ac{aus} while in \ac{gb} it is concentrated onto a few units.  In this context, it has been shown in the literature that the former approach leads to a better frequency performance \cite{9495014}.  Another key reason why \ac{aus} shows a better frequency regulation performance is that \ac{aus} utilizes and \ac{agc} whereas \ac{gb} does not.

\begin{figure}[htb]
  \subfigure[\ac{aus}: May 2019]{\resizebox{0.495\linewidth}{!}{\includegraphics{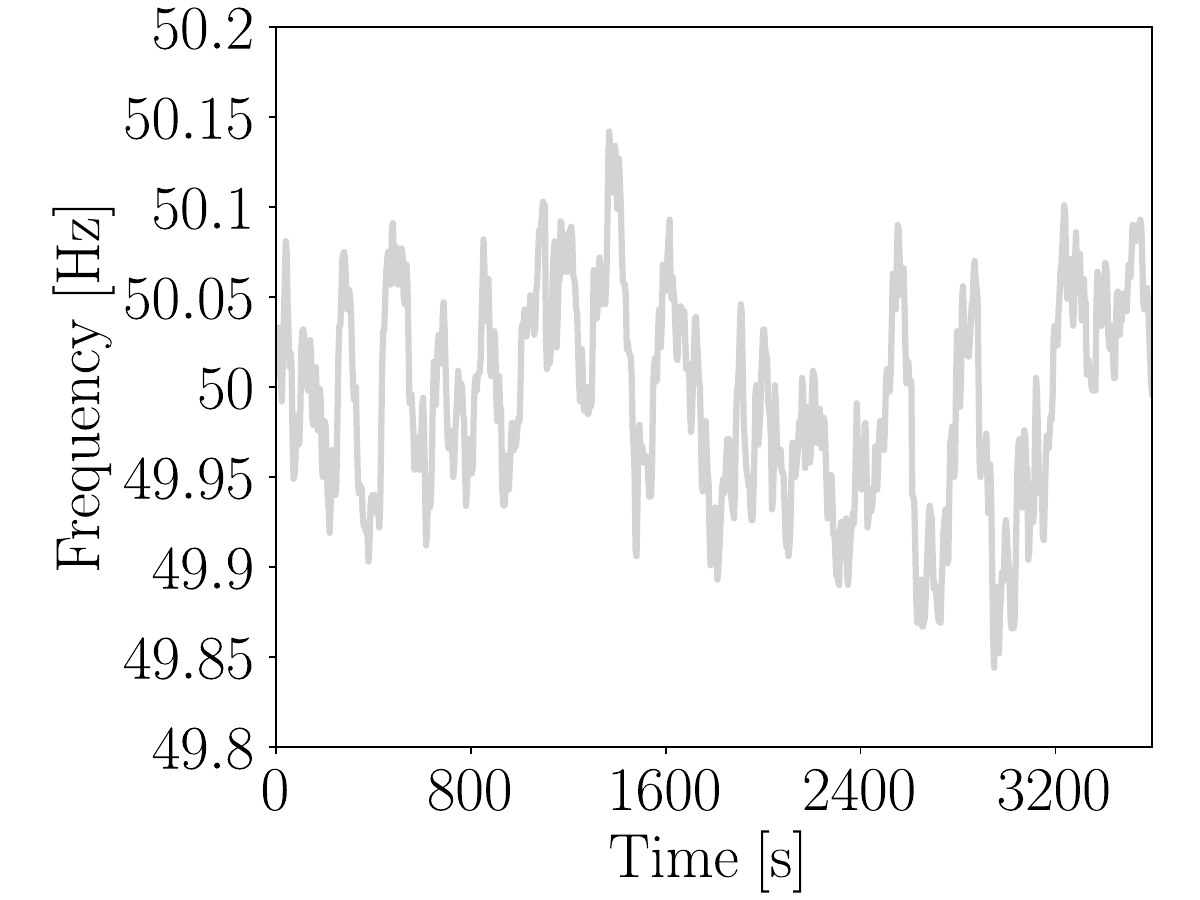}}}
  \subfigure[\ac{aus}: May 2024]{\resizebox{0.495\linewidth}{!}{\includegraphics{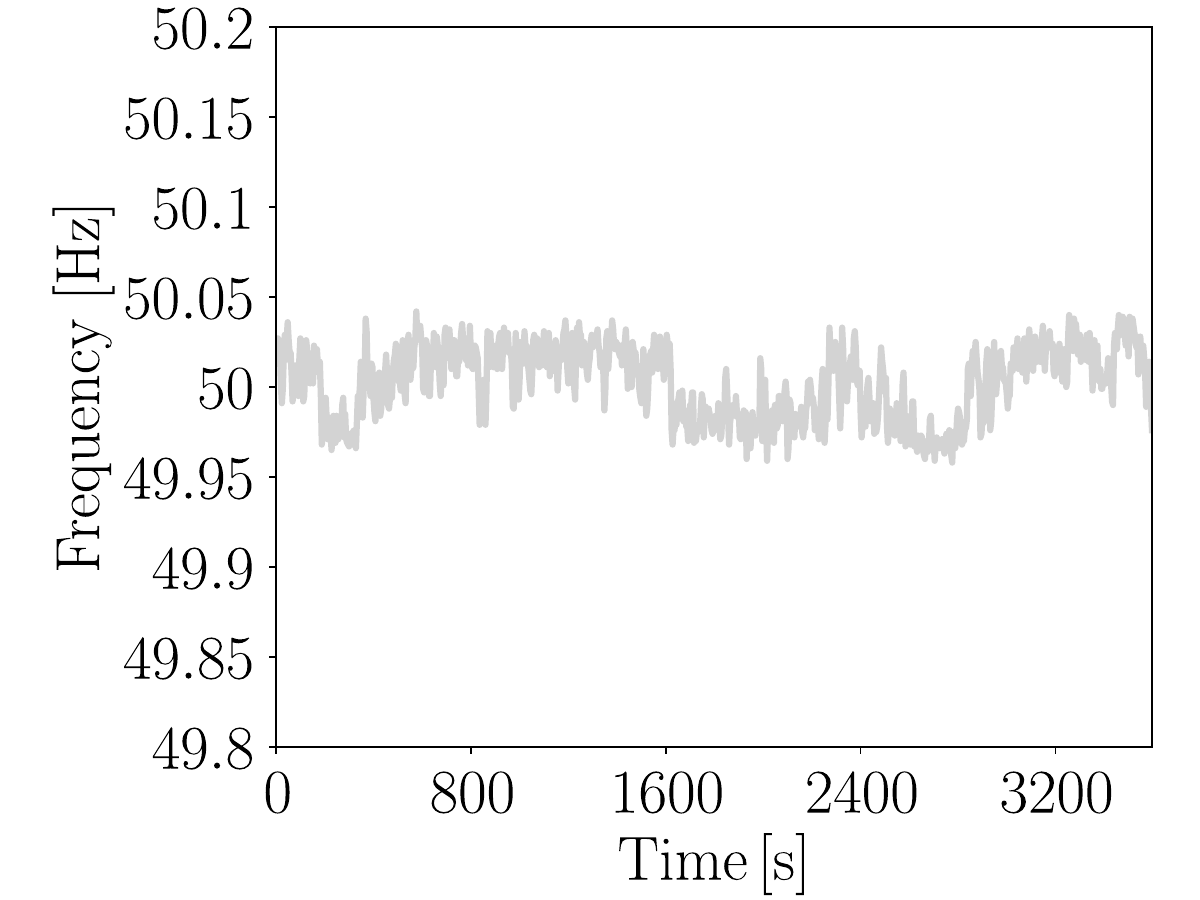}}}
  \subfigure[\ac{gb}: May 2019]{\resizebox{0.495\linewidth}{!}{\includegraphics{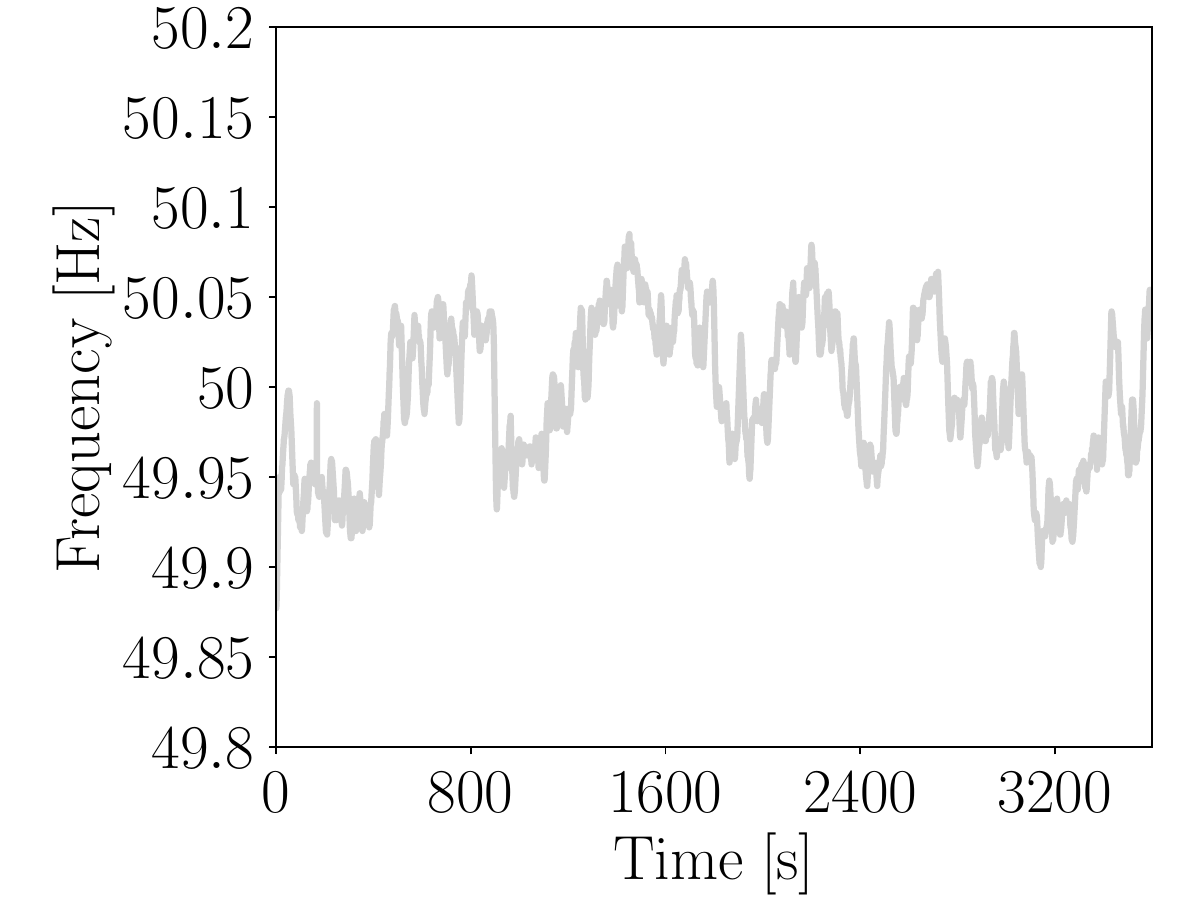}}}
  \subfigure[\ac{gb}: May 2024]{\resizebox{0.495\linewidth}{!}{\includegraphics{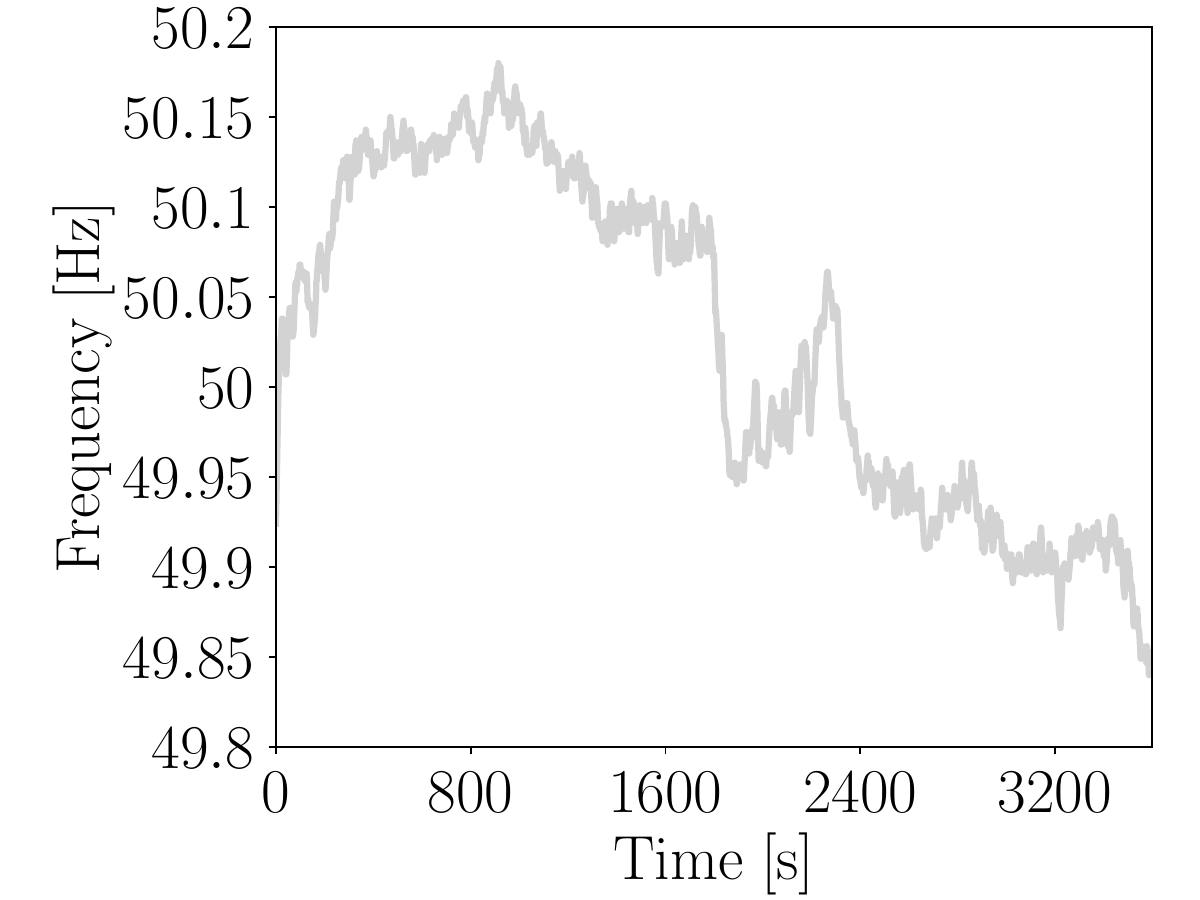}}} \\
  \caption[Comparison of frequency traces of GB and AUS]{Comparison of frequency traces of \ac{gb} and \ac{aus} power systems for 1 hour on May 21, 2019 and 2024, respectively.}
  \label{fig:05:compmay}
\end{figure}

The \ac{aus} system, in fact, was showing a decline in frequency control performance under normal conditions (similar to \ac{gb}) before 2020 where frequency regulation was predominantly managed through \ac{agc} and relatively wide dead-band (for example, $\pm 150$ mHz).  However, managing frequency regulation only through \ac{agc} seemed a hard task for \ac{aemo}.  For this reason, in 2020 the Australian Energy Market Commission introduced mandatory \ac{pfc} rule for all generators, including from \acp{ibr} such as wind power but excluding \acp{der} such as roof-top \ac{pv} \cite{aemo2}.  The mandatory \ac{pfc} provision with narrow dead-band ($\pm 15$ mHz) and proportional droop response led to a significant improvement of frequency performance in \ac{aus}.  This is illustrated in Figure \ref{fig:05:compmay} where frequency traces of the \ac{gb} and \ac{aus} systems are compared for the same hour and day (May 21, 2019 and 2024).  Frequency variations in \ac{aus} have dramatically improved compared to 2019 but that is not the case in \ac{gb}.  This suggests that a potential solution for \ac{gb} is to impose \ac{pfc} with narrow dead-band, even though this might be difficult to implement in practice by all generators as narrow dead-band increases wear and tear.  For instance, large dead-bands may be preferred for nuclear units to avoid movement in active power output from scheduled values caused by changes in frequency \cite{8274729}.

If the mandatory \ac{pfc} rule with narrow dead-band is not possible, then implementing an \ac{agc} (well-proven frequency regulation capability) should be another viable solution for \ac{gb} to consider.\footnote{For instance, the Nordic \acp{tso} identified \ac{afrr}/\ac{agc} as one of the main measures to stop the weakening
trend of the frequency quality and introduced it back in 2013 \cite{nordicphilosophy}.}  We believe this is important considering the fact that the \ac{ic} ramp rates in \ac{gb} are significantly higher than in \ac{aips} (100 MW/min vs 5 MW/min).  In other words, several \acp{ic} ramping at the same time exacerbates the control of system frequency \cite{NGESOIC} and, thus, an \ac{agc} may be best to deal with it.  In addition, despite \ac{pfc} being a fast-acting and continuous method of control, it is not perfect tracking and thus will not bring the frequency error to zero in steady-state in contrast to \ac{agc} (includes an integrator term) \cite{9361269}. Another solution to better manage the real-time power imbalance could be fast generation dispatch but it has been shown in the literature that \ac{agc} outperforms it in terms of frequency regulation performance \cite{10002304}.  The \ac{gb} might also consider increasing the volumes of \ac{dr} and \ac{dm} products.  As a matter of fact, in February 2025, \ac{neso} increased these volumes to better manage significant MW movements observed in recent weeks, see Table \ref{tab:05:drdmincrease} \cite{NGESOnew}.  However, these new volume requirements will lead to additional costs.

\begin{table}[htb]
  \centering
  \caption{DR \& DM requirements increase.} 
  \label{tab:05:drdmincrease}
  \resizebox{1.0\linewidth}{!}{
  \begin{tabular}{lcccc}
    \hline
    & \ac{dr}-Low & \ac{dr}-High & \ac{dm}-Low & \ac{dm}-High \\
    & [MW] & [MW] & [MW] & [MW]  \\
    \hline
    Before \; February 2, 2025 & 330 & 330 & 170 & 200 \\
    Since \; February 2, 2025 & 480 & 480 & 300 & 300 \\
    \hline
  \end{tabular}}
\end{table}

\subsubsection{${\rm \Delta} f$-based comparison}

\begin{figure}[htb]
  \subfigure[$i = 4$ s]{\resizebox{0.495\linewidth}{!}{\includegraphics{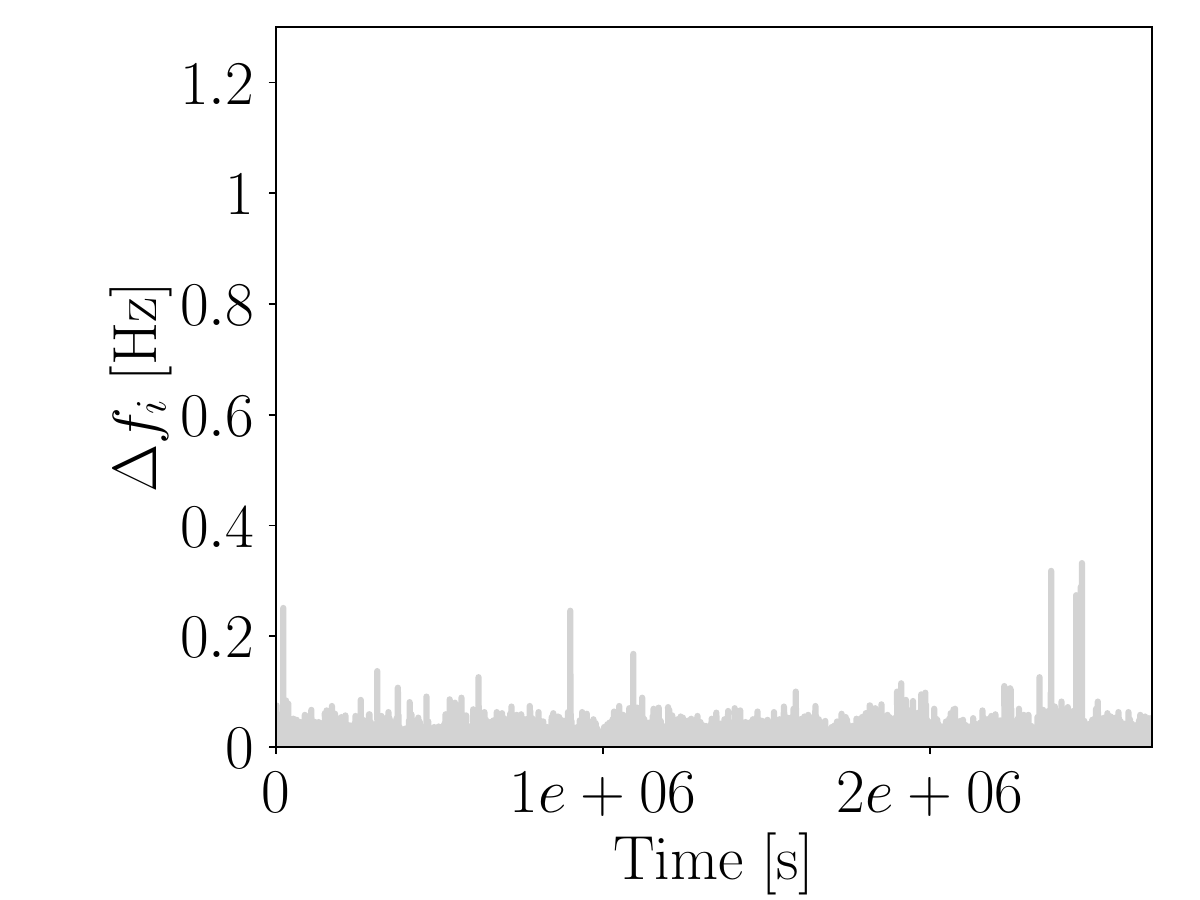}}}
  \subfigure[$i = 60$ s]{\resizebox{0.495\linewidth}{!}{\includegraphics{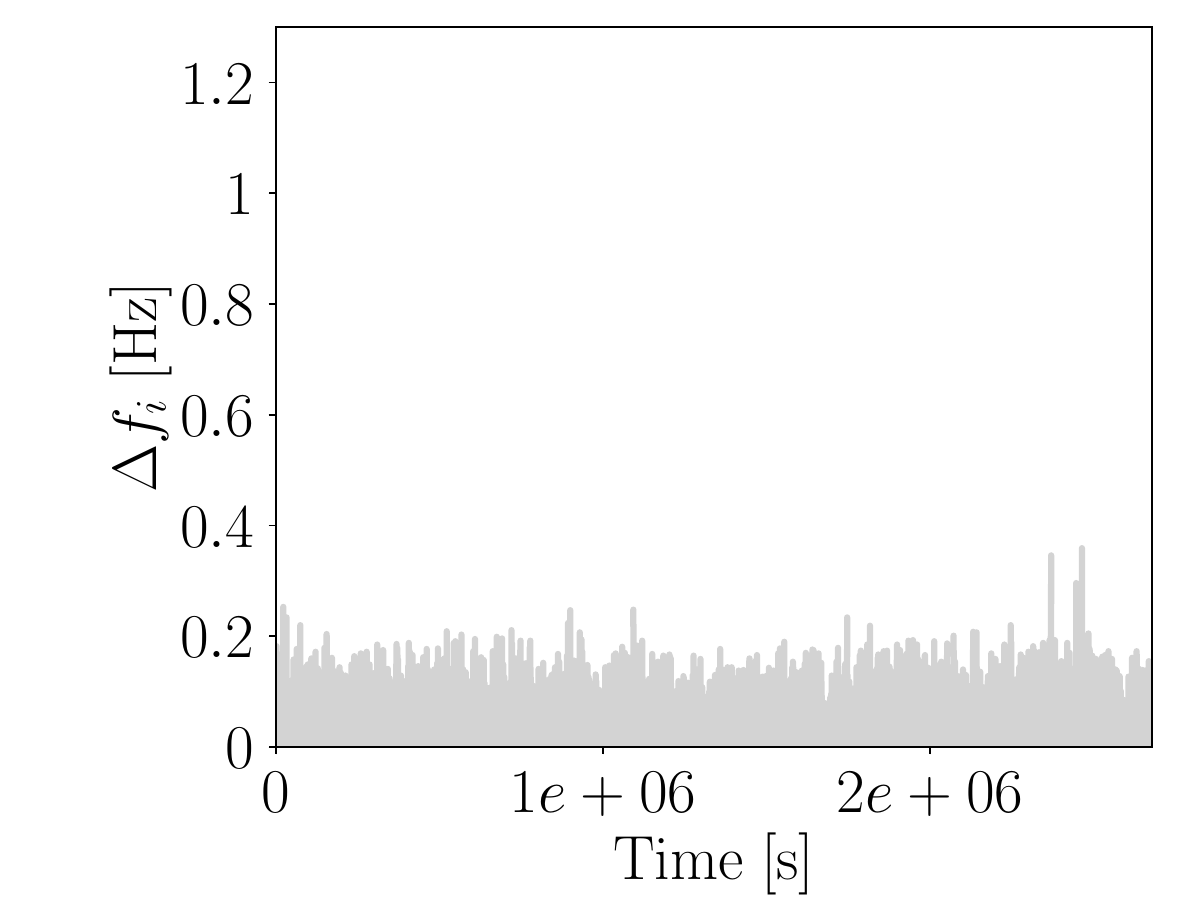}}}
  \subfigure[$i = 300$ s]{\resizebox{0.495\linewidth}{!}{\includegraphics{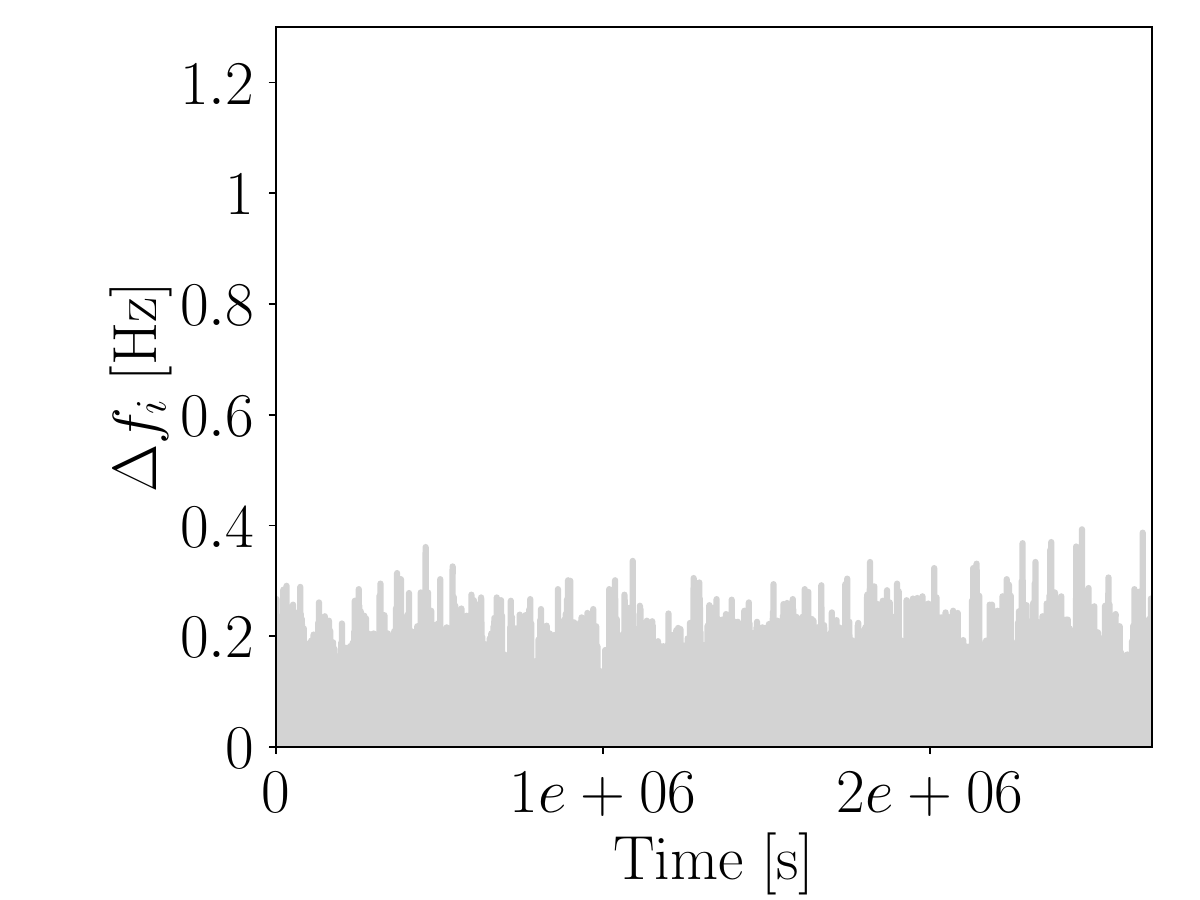}}}
  \subfigure[$i = 900$ s]{\resizebox{0.495\linewidth}{!}{\includegraphics{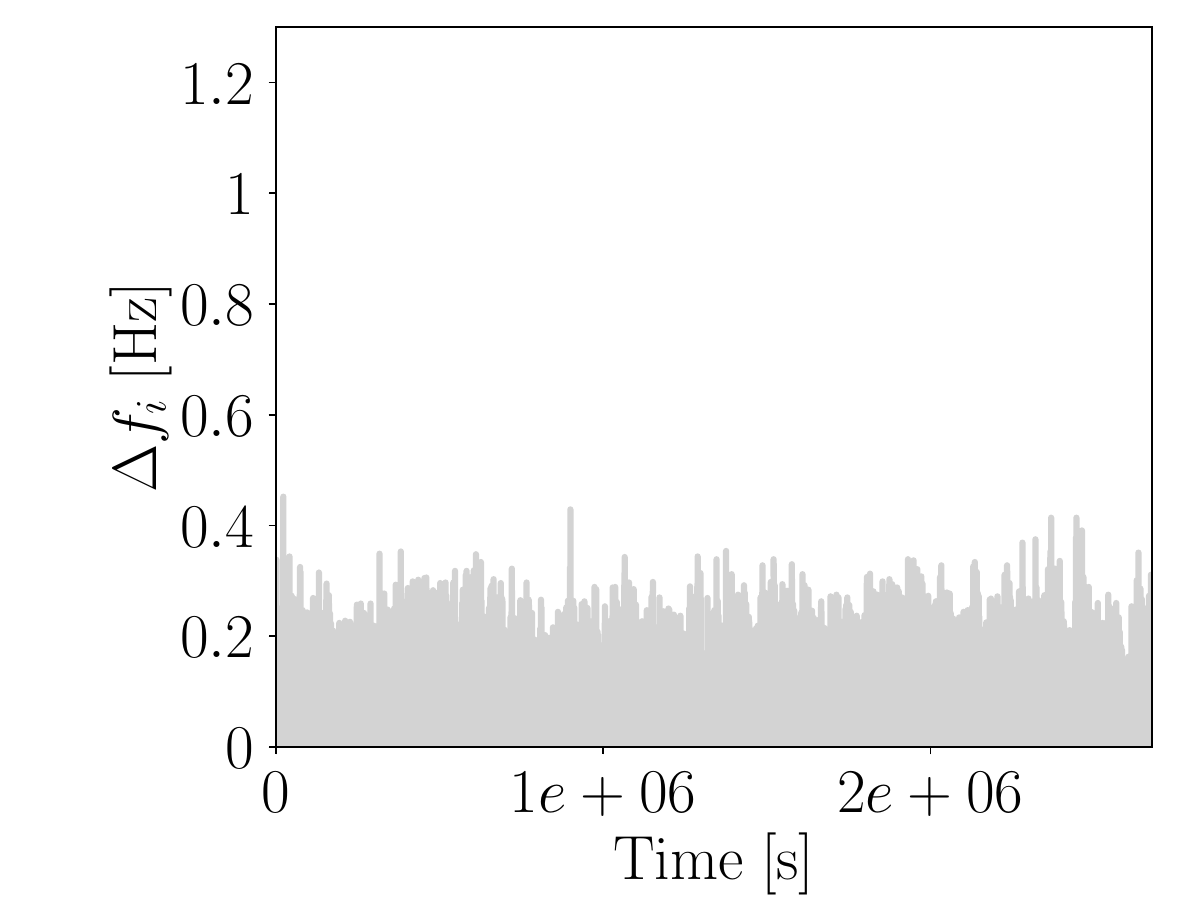}}} \\
  \caption[$\Dfi$ variations in the GB power system]{$\Dfi$ variations in the \ac{gb} power system. 
  }
  \label{fig:05:gbrocof}
\end{figure}

\begin{figure}[htb]
  \subfigure[$i = 4$ s]{\resizebox{0.495\linewidth}{!}{\includegraphics{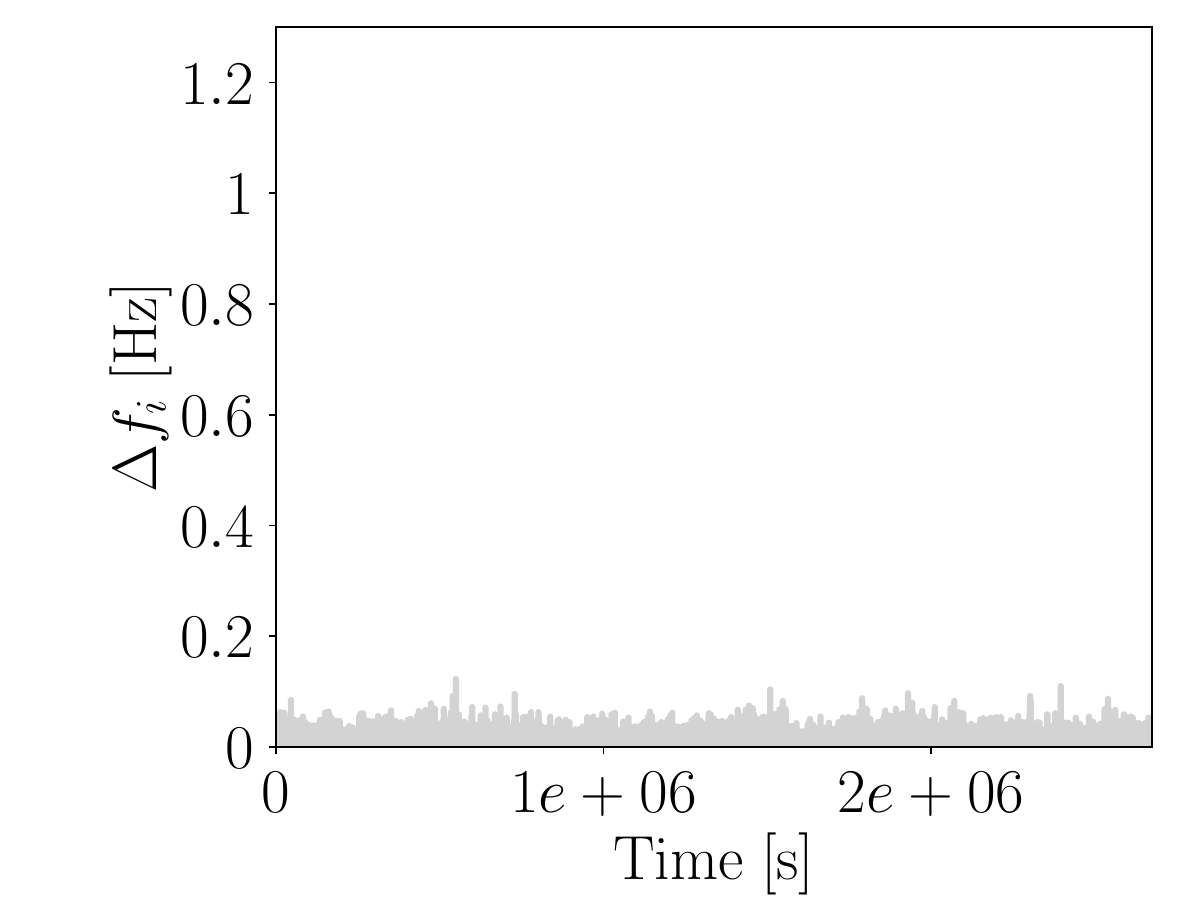}}}
  \subfigure[$i = 60$ s]{\resizebox{0.495\linewidth}{!}{\includegraphics{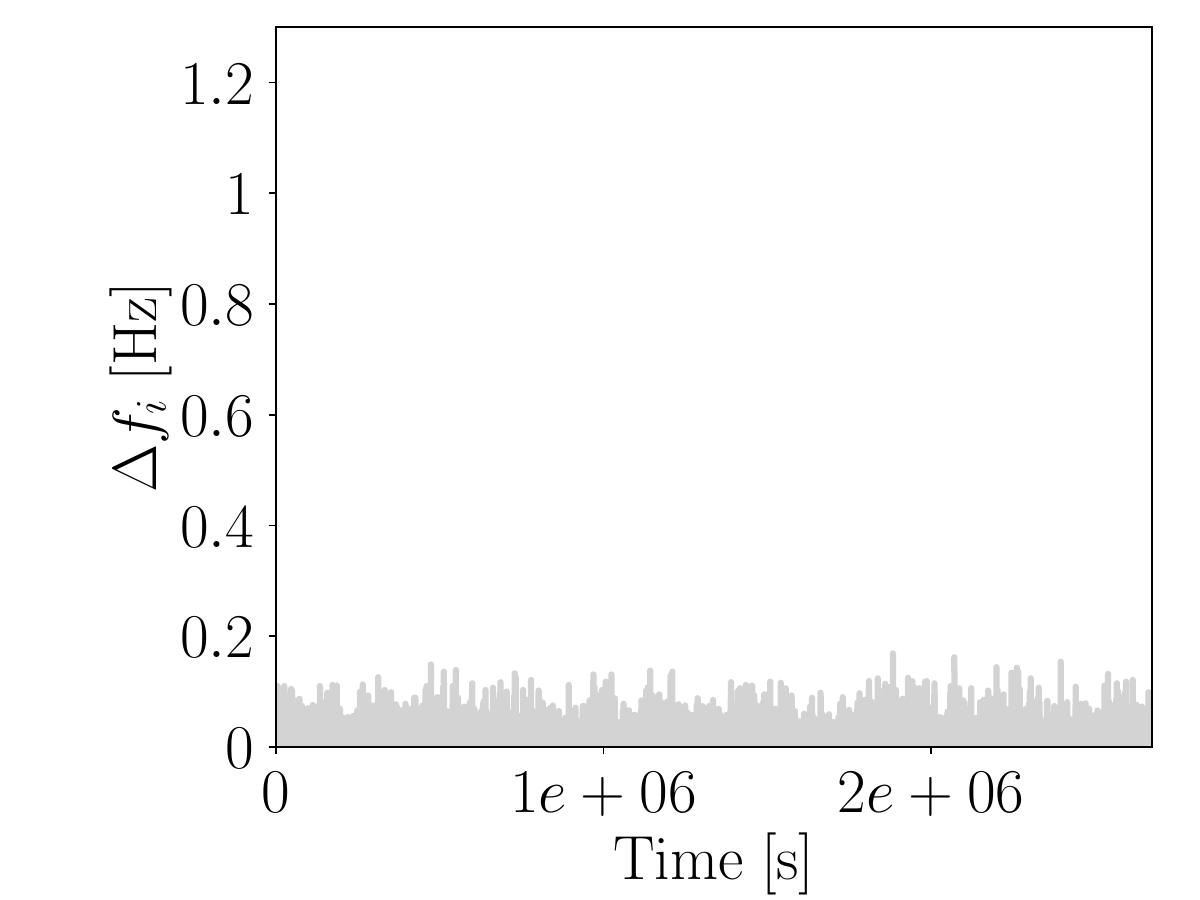}}}
  \subfigure[$i = 300$ s]{\resizebox{0.495\linewidth}{!}{\includegraphics{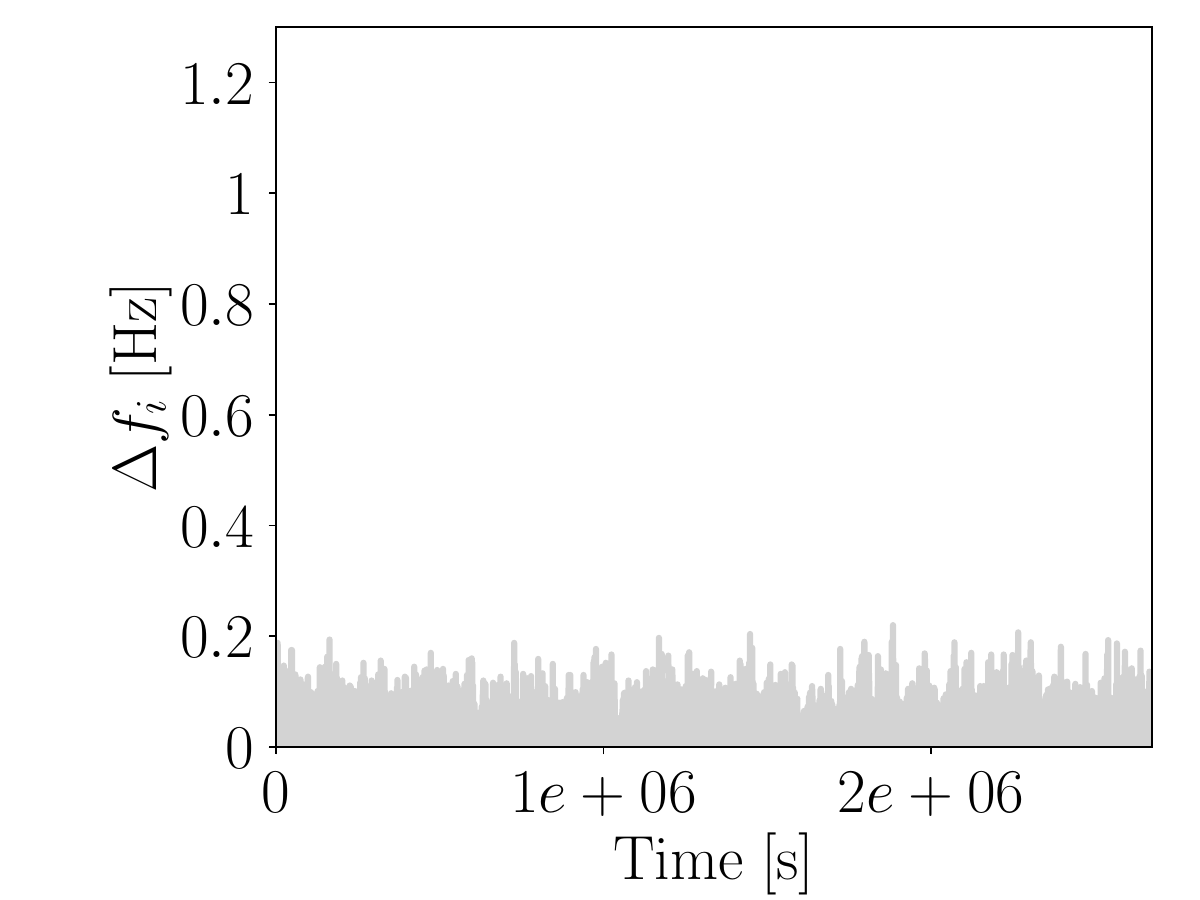}}}
  \subfigure[$i = 900$ s]{\resizebox{0.495\linewidth}{!}{\includegraphics{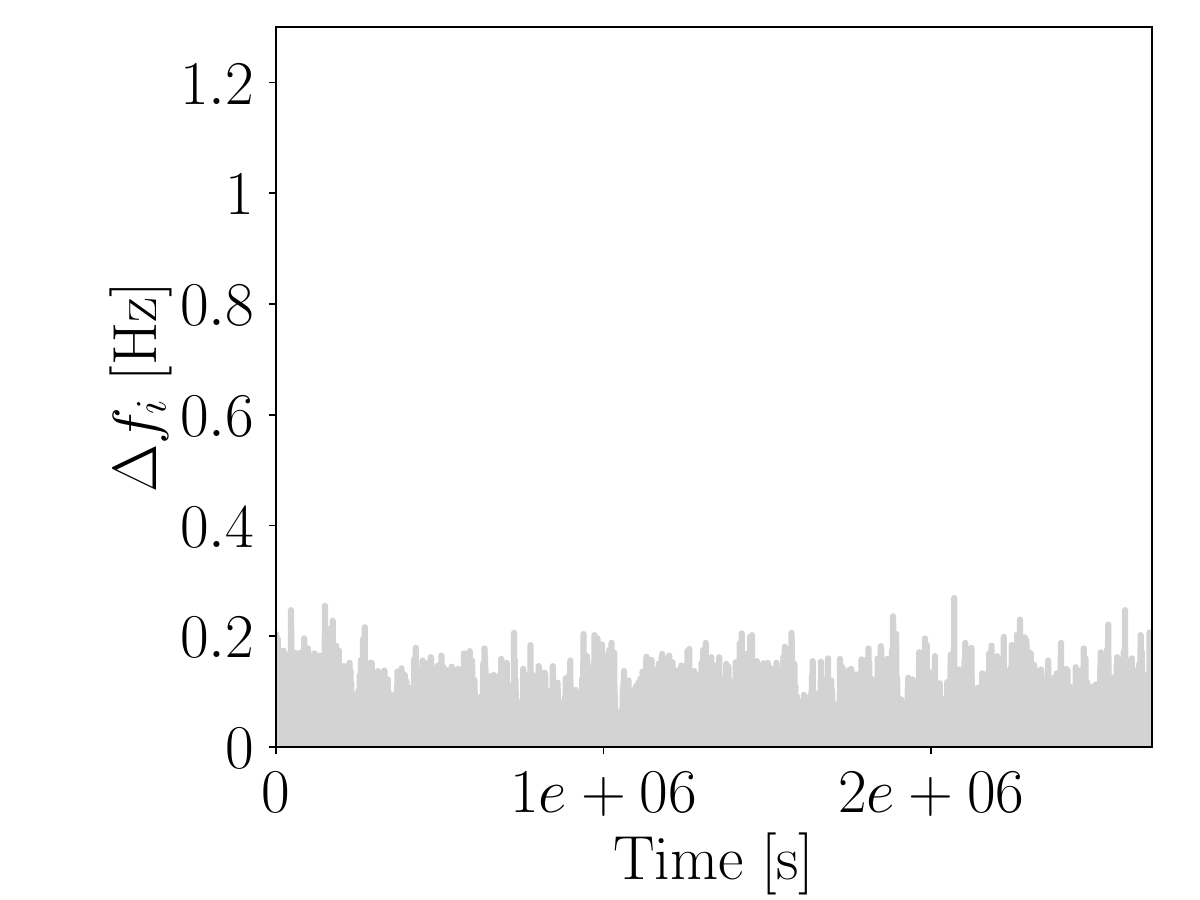}}} \\
  \caption[$\Dfi$ variations in the AIPS]{$\Dfi$ variations in the \ac{aips}.} 
  \label{fig:05:irishroc}
\end{figure}

\begin{figure}[htb]
  \subfigure[$i = 4$ s]{\resizebox{0.495\linewidth}{!}{\includegraphics{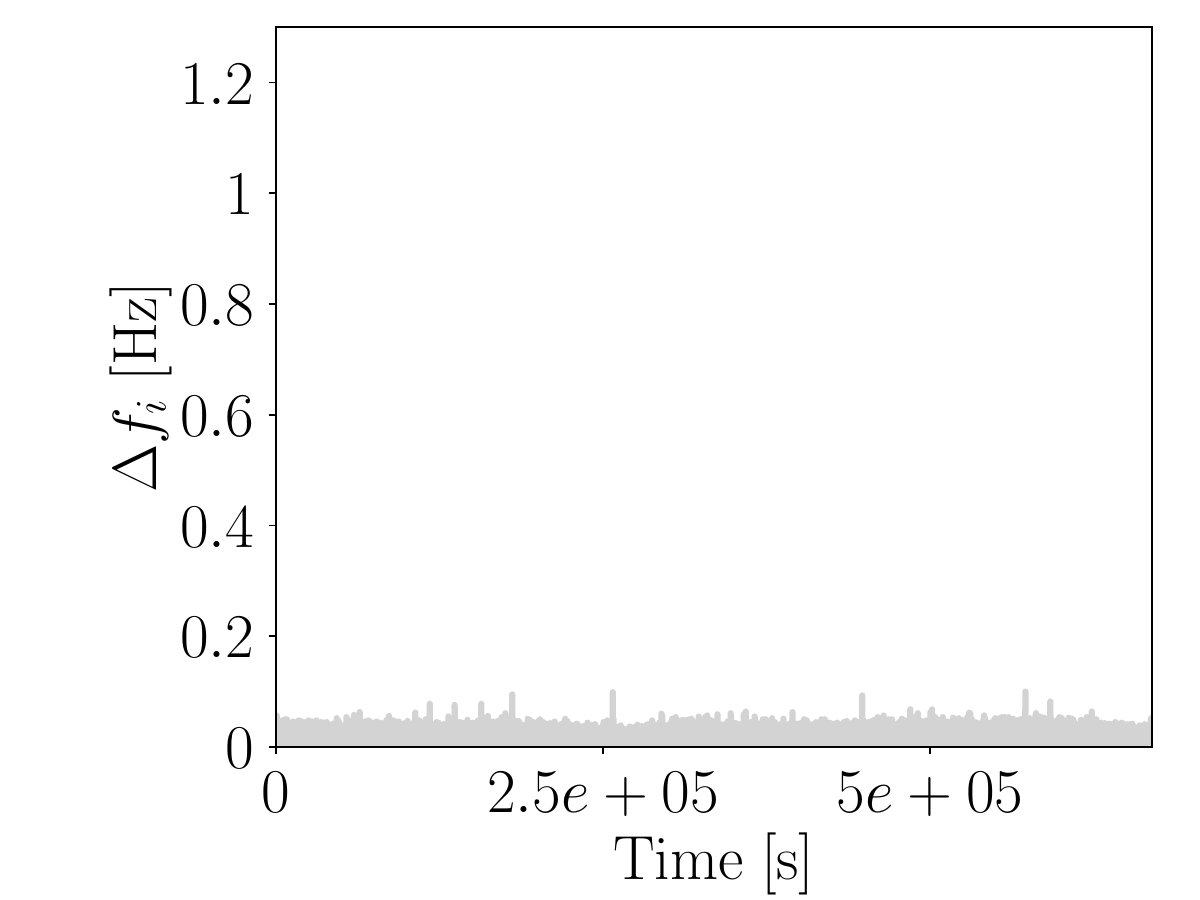}}}
  \subfigure[$i = 60$ s]{\resizebox{0.495\linewidth}{!}{\includegraphics{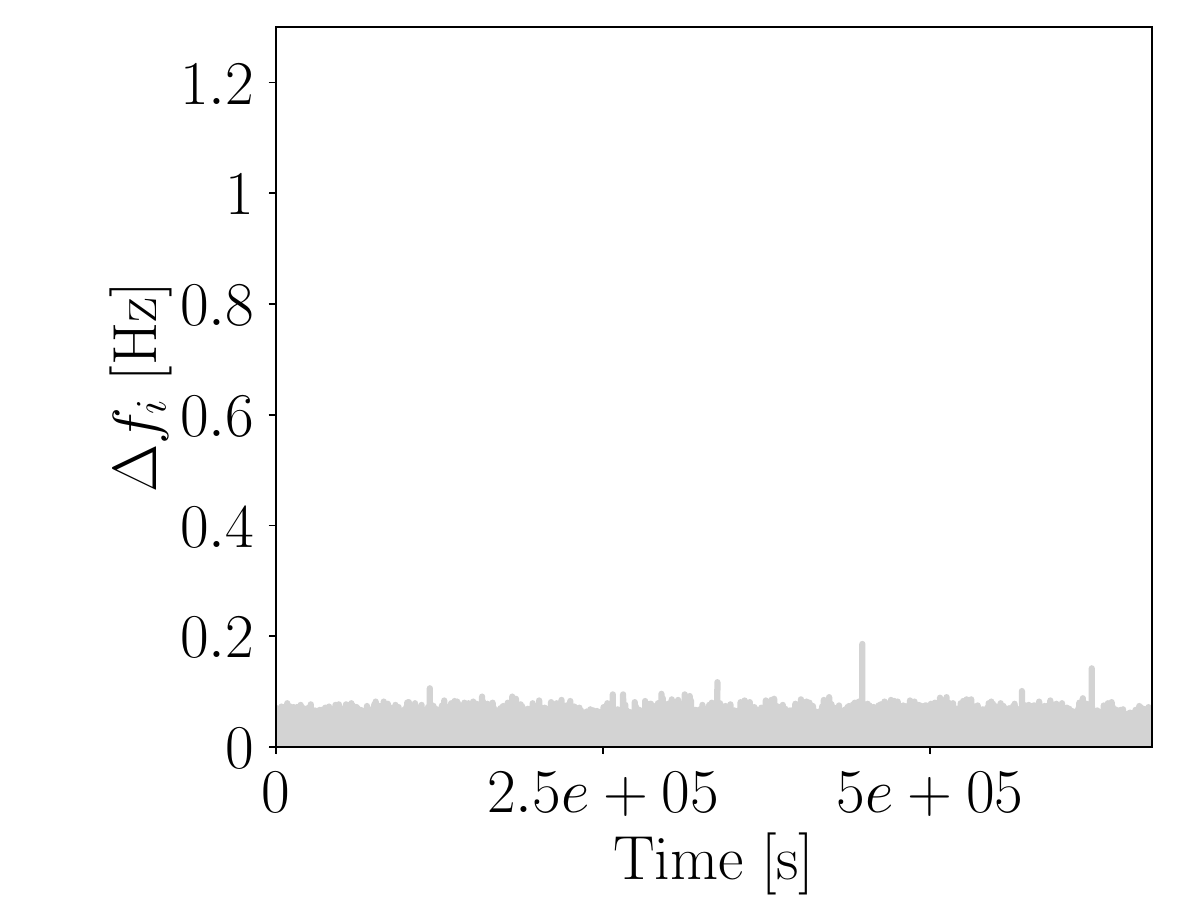}}}
  \subfigure[$i = 300$ s]{\resizebox{0.495\linewidth}{!}{\includegraphics{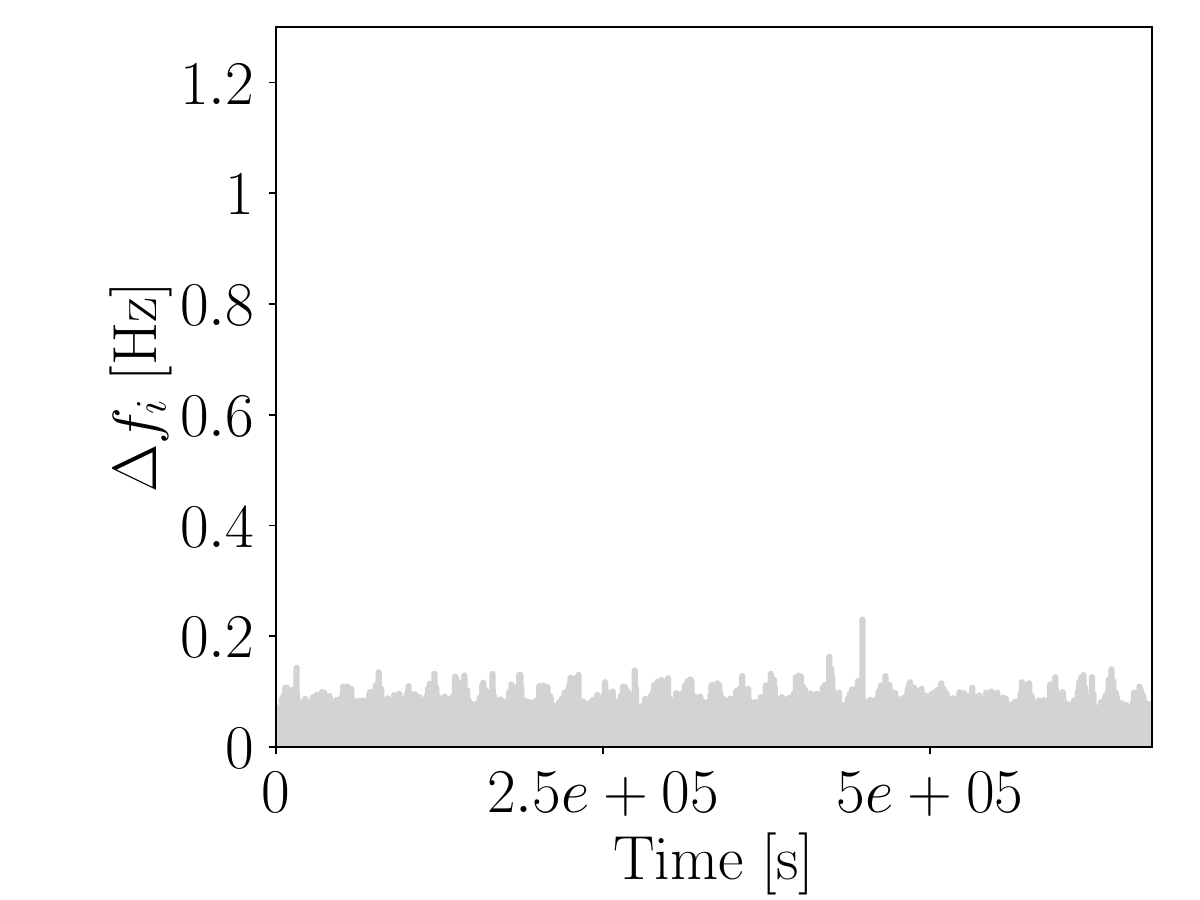}}}
  \subfigure[$i = 900$ s]{\resizebox{0.495\linewidth}{!}{\includegraphics{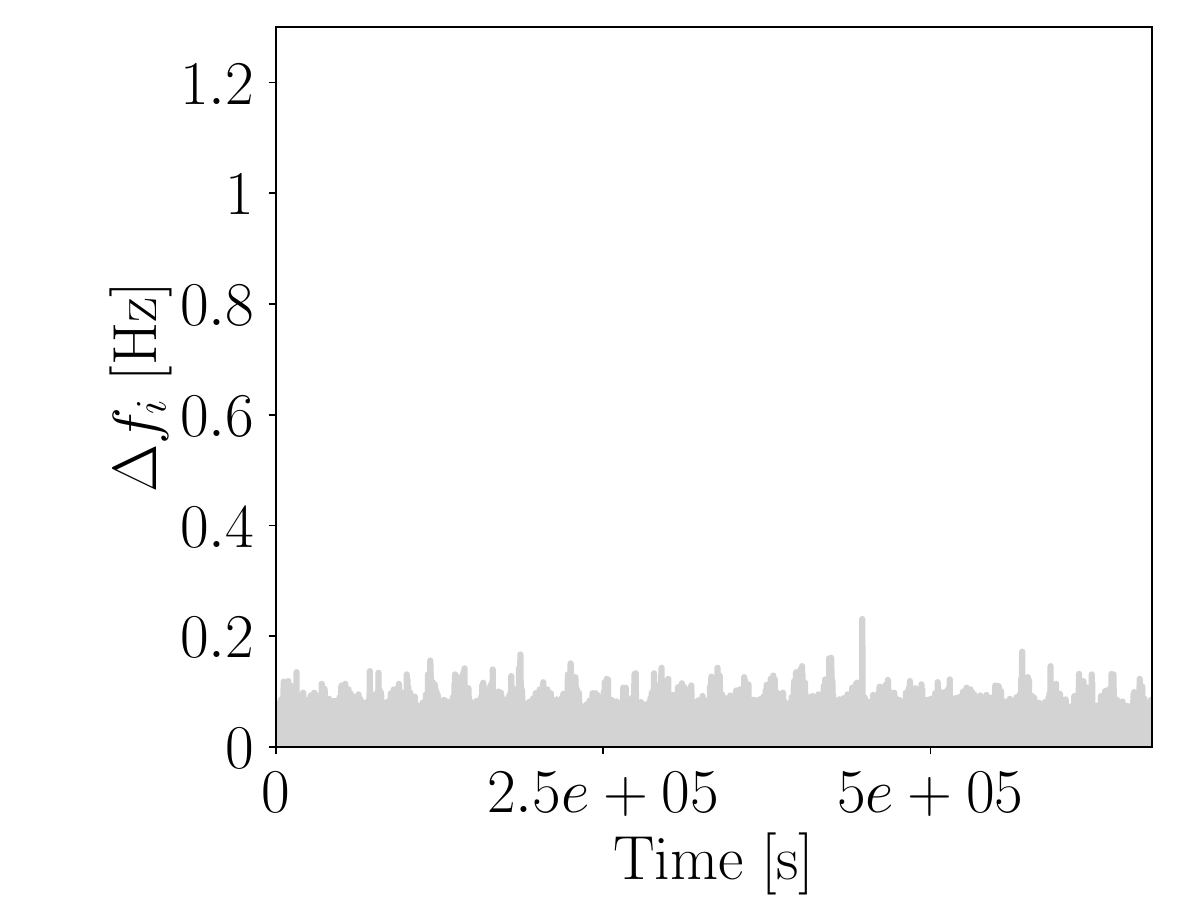}}} \\
  \caption[$\Dfi$ variations in the AUS power system]{$\Dfi$ variations in the \ac{aus} power system. 
  }
  \label{fig:05:mainhrocof}
\end{figure}

\begin{figure}[htb]
  \subfigure[$i = 4$ s]{\resizebox{0.495\linewidth}{!}{\includegraphics{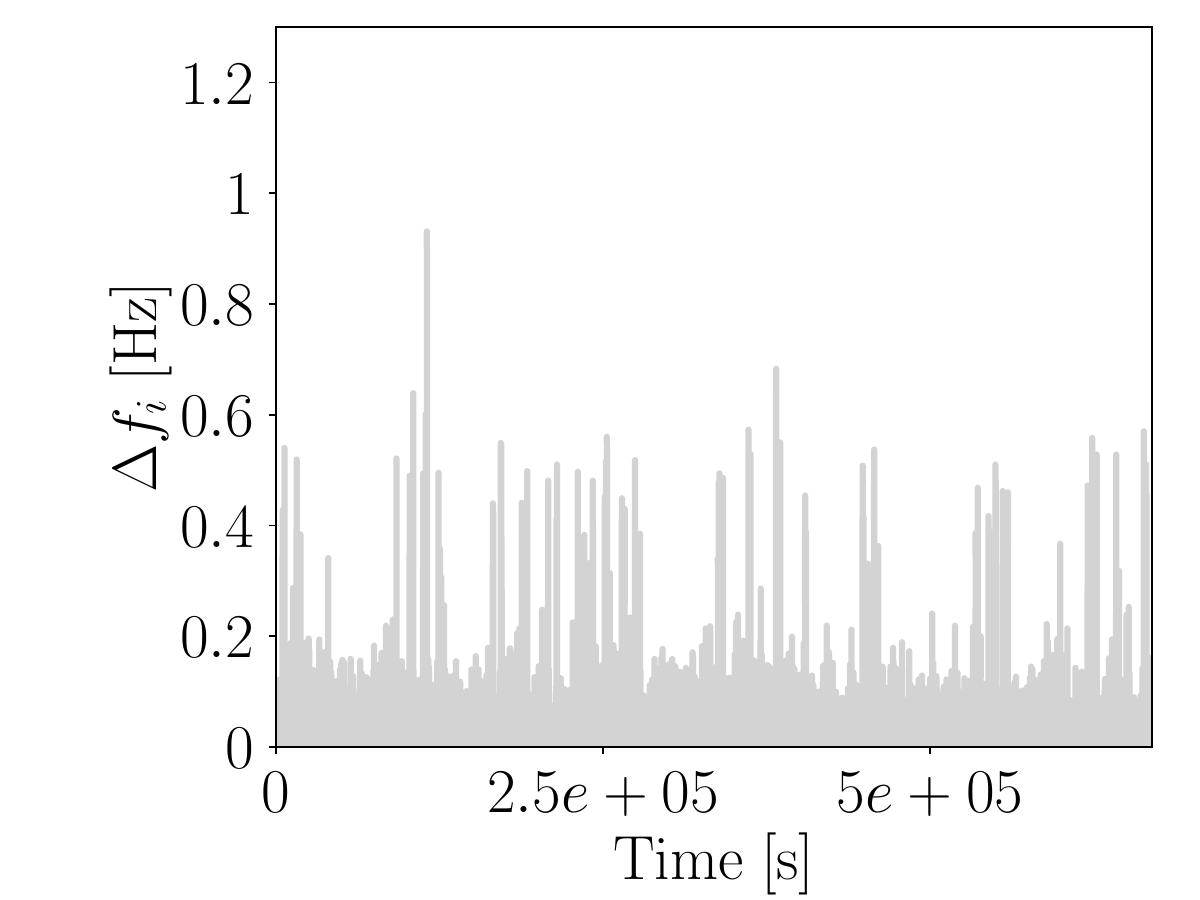}}}
  \subfigure[$i = 60$ s]{\resizebox{0.495\linewidth}{!}{\includegraphics{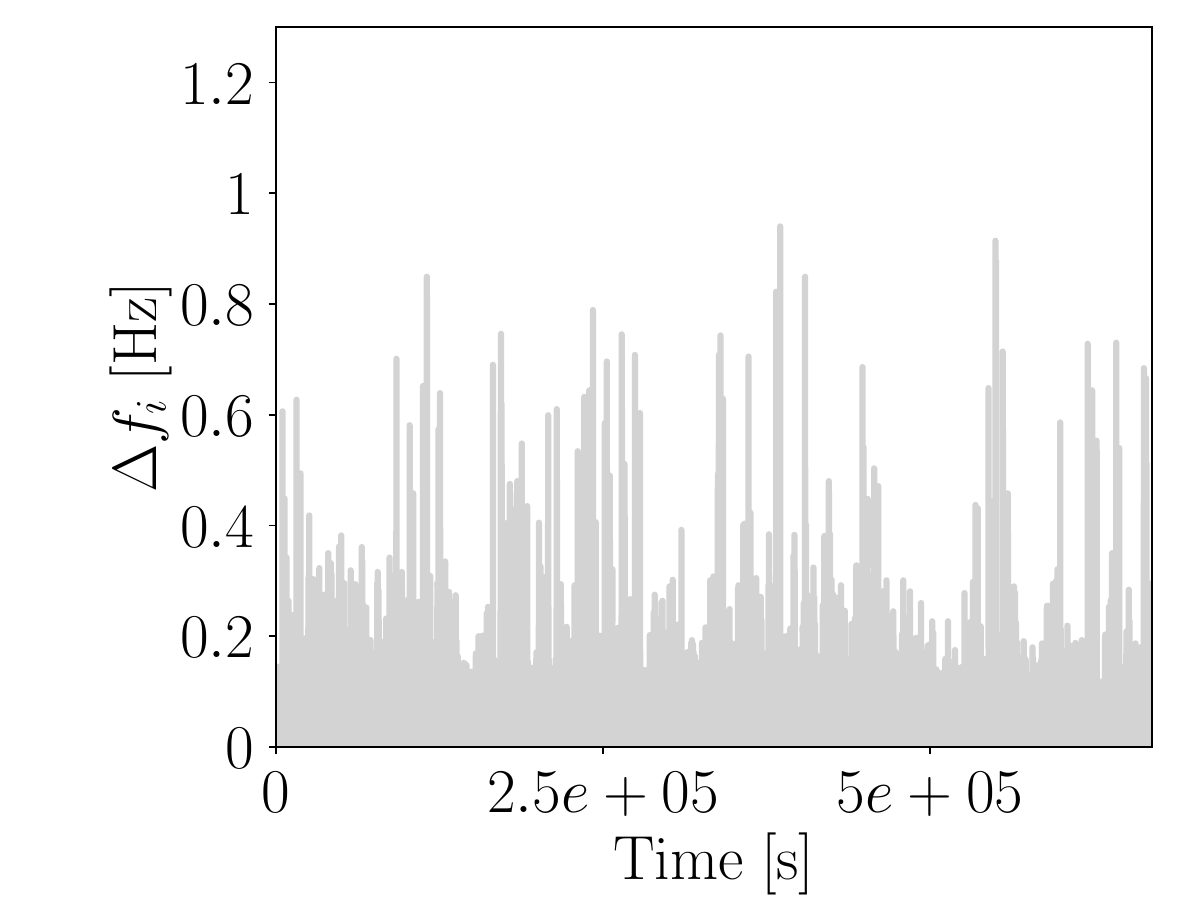}}}
  \subfigure[$i = 300$ s]{\resizebox{0.495\linewidth}{!}{\includegraphics{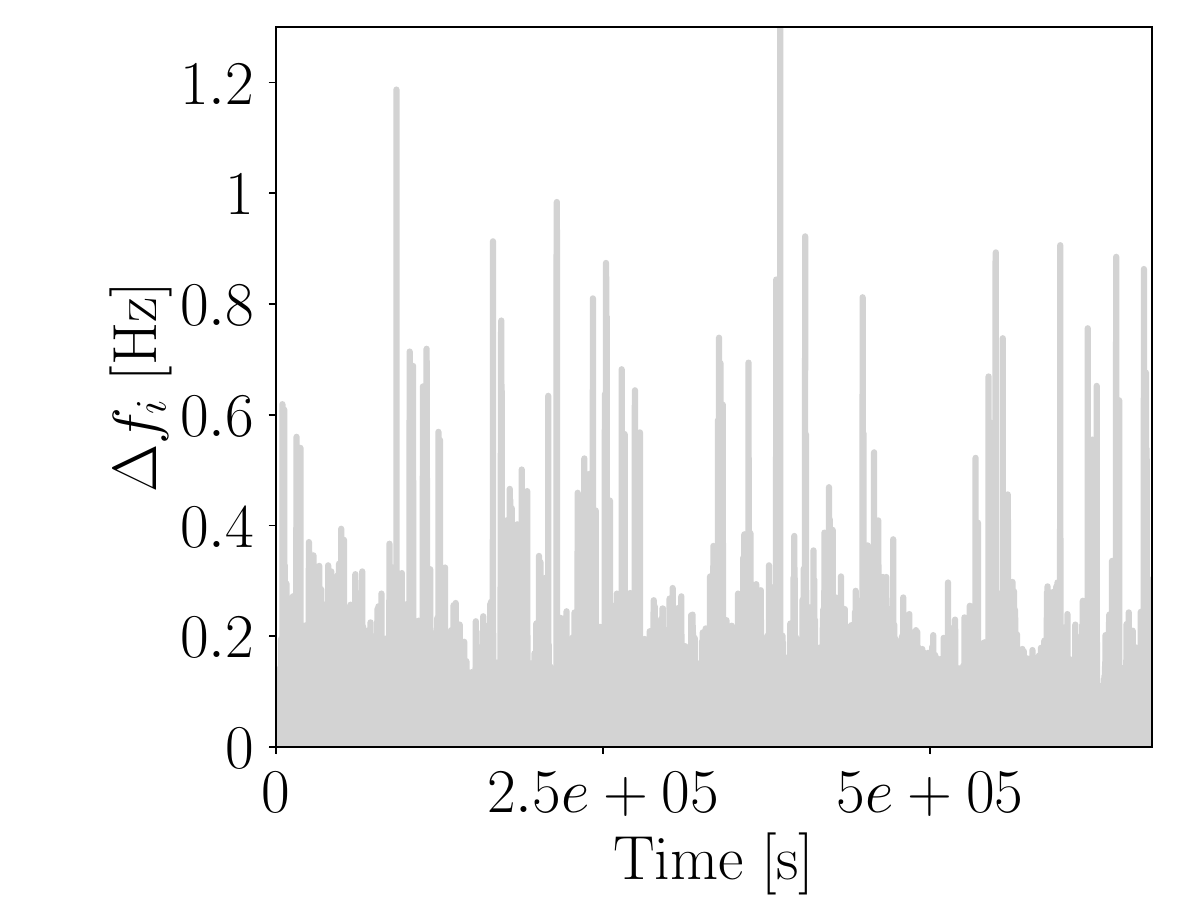}}}
  \subfigure[$i = 900$ s]{\resizebox{0.495\linewidth}{!}{\includegraphics{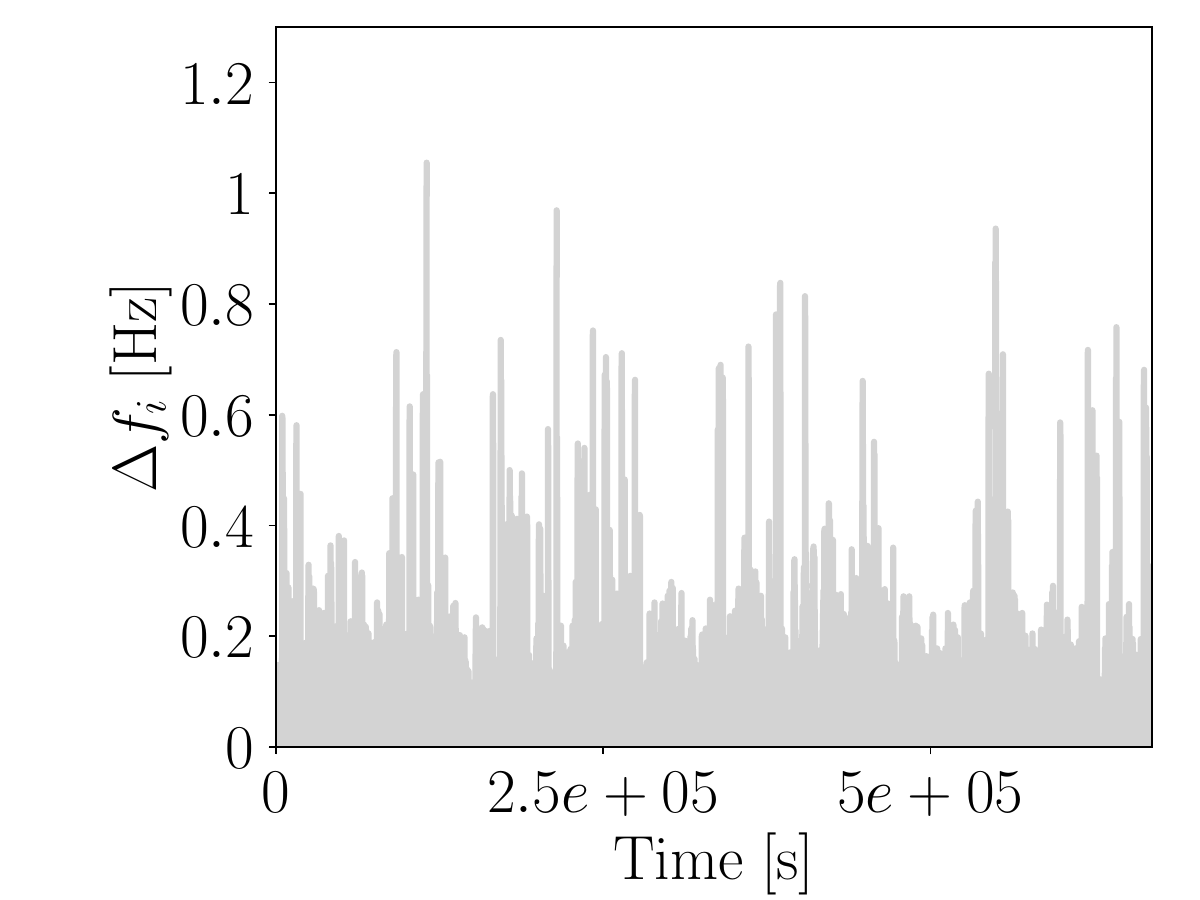}}} \\
  \caption[$\Dfi$ variations in the TAS power system]{$\Dfi$ variations in the \ac{tas} power system. 
  }
  \label{fig:05:tasrocof}
\end{figure}

To get a better insight into the dynamics of frequency variations under normal conditions of the four power systems, we calculate $\Dfi$ over different time periods.  With this aim, we select March 2024 as the reference month.  Specifically, Figures \ref{fig:05:gbrocof}, \ref{fig:05:irishroc}, \ref{fig:05:mainhrocof} and \ref{fig:05:tasrocof} show $\Dfi$ calculated for $i=4$ s, $i=60$ s, $i=300$ s and $i=900$ s time periods.  As expected, considering its small size, \ac{tas} system shows the highest $\Dfi$ across all the considered time periods compared to \ac{gb}, \ac{aips} and \ac{aus} systems.

\ac{gb} follows \ac{tas} with the second highest $\Dfi$ experienced across 4 s, 60 s, 300 s and 900 s, respectively.  While this is a counter intuitive result from the system size point of view, it is somehow expected following the frequency quality deterioration of \ac{gb} discussed in the previous section.  Figure \ref{fig:05:gbrocof} shows another insightful result, that is, it appears that $\Dfi$ gets higher for higher time periods, for example, ${\rm \Delta} f_{300}$ higher than ${\rm \Delta} f_{60}$.  These results are consistent with the observation from \ac{neso} that states that during 2021 around 65\% of the total time outside $\pm 100$ mHz is due to events lasting 60 s or more, and only around 15\% of the total time outside $\pm 100$ mHz is due to events lasting 5 minutes or more \cite{NGESO}.  This observation is not as obvious for \ac{aips} and \ac{aus} systems.  This means that events that lead to frequency variations in \ac{aips} and \ac{aus} last less time and are compensated more quickly than in \ac{gb}.

While one might argue that $\Dfi$ in the \ac{aus} system are lower compared to \ac{gb} due to \ac{agc} in \ac{aus}, that is not the case when comparing \ac{gb} with \ac{aips} system.  Specifically, since both \ac{aips} and \ac{gb} systems have a manual \ac{agc}, then the main difference between the two systems is the \ac{pfc} rule, that is, mandatory in \ac{aips} and non-mandatory in \ac{gb}.  Another potential reason (difference) why frequency variations last more in \ac{gb} could be related to market-driven variations that might be significantly higher in \ac{gb} due to, for example, higher \ac{ic} ramp rates and more demand volatility due to GW level flexibility services being procured by \ac{neso} and \acp{dso} in \ac{gb} \cite{10407387}.  For example, reference \cite{HOMAN2021116723} suggest that one of the causes of frequency events in \ac{gb} is a high rate of change of demand.  In fact, if one looks at the flexibility service volumes contracted in \ac{gb} in Table \ref{tab:05:flexvolumes}, in particular, post-fault products (``Dynamic'' and ``Restore'') that have response times in time scale of minutes, it might be concluded that demand response has increased significantly in recent years \cite{epriflex}.

\begin{table*}[htb]
  \centering
  \caption[Service products contracted across all GB utilities]{Evolution of service products contracted across all \ac{gb} utilities.} 
  \label{tab:05:flexvolumes}
  \begin{tabular}{lccccc}
    \hline
    Flexibility Service  & 2018 & 2019 & 2020 & 2021 & 2021/2022 \\
    & [MW, \%] & [MW, \%] & [MW, \%] & [MW, \%] & [MW, \%] \\
    \hline
    Sustain & 0, 0 & 0, 0 & 2, 0 & 13.3, 1 & 28.1, 2\\
    Secure & 23.8, 20 & 10.3, 4 & 105.1, 9 & 262.6, 16 & 375.2, 20\\
    Dynamic & 33.8, 29 & 120.8, 47 & 555.9, 48 & 729.7, 45 & 925.7, 50\\
    Restore & 58.5, 50 & 125.1, 49 & 502.5, 43 & 603, 37 & 538, 29\\
    Total & 116.1, 100 & 256.2, 100 & 1165.5, 100 & 1608.5, 100 & 1867, 100\\
    \hline
  \end{tabular}
\end{table*}

\subsubsection{Frequency standard deviation-based comparison}
\label{sec:05:stdresults}

\begin{figure}[thb!]
  \begin{center}
    \resizebox{1.0\linewidth}{!}{\includegraphics{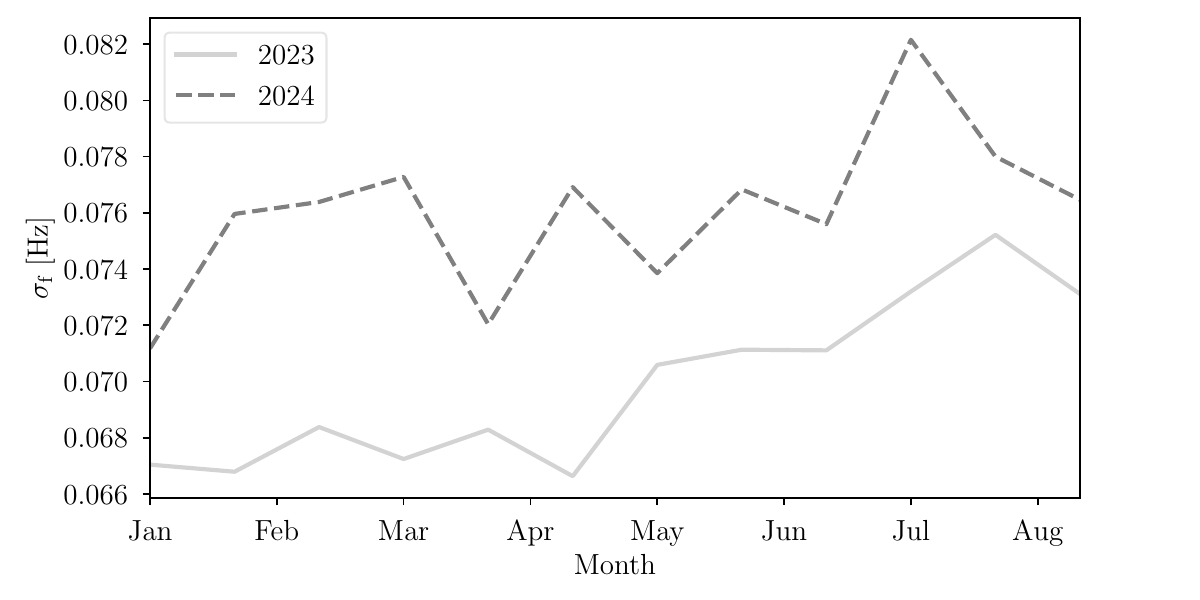}}
    \caption[Evolution of $\sigmaf$ in GB in the last two years]{Evolution of $\sigmaf$ in the \ac{gb} power system in 2023 and 2024.}
    \label{fig:05:std_gb}
  \end{center}
\end{figure}

We conclude the frequency control strength comparison for normal system conditions by comparing the various $\sigma$-based metrics presented in Section \ref{sec:05:std}.  Results are presented in Table \ref{tab:05:results}.  Similar to the frequency quality and $\Dfi$-based results, \ac{gb} has the highest $\sigmaf$ (0.076 Hz) compared to \ac{aips} (0.042 Hz), \ac{aus} (0.025 Hz), and \ac{tas} (0.042 Hz) systems.  These different $\sigmaf$ in March 2024 match quite well the $\sigmaf$ in 2023 for the four power systems.  Specifically, the $\sigmaf$ in 2023 for \ac{gb}, \ac{aips}, \ac{aus} and \ac{tas} are 0.069 Hz, 0.042 Hz, 0.025 Hz, and 0.040 Hz, respectively.  These results support the idea that one month frequency analysis/data (March 2024) is representative enough of the overall frequency performance of the selected power systems.  In fact, $\sigmaf$ has recently constantly increased in the \ac{gb} power system as shown in Figure \ref{fig:05:std_gb}.  As mentioned above, because of this increased frequency volatility, \ac{neso} increased the volumes of procured \ac{pfr} reserves namely \ac{dm} and \ac{dr} reserve volumes.

\begin{table}[t!]
  \centering
  \caption{Summary of standard deviation-based results in March 2024.}
  \label{tab:05:results}
  \begin{tabular}{cccccc}
    \hline
    Power System  & $\sigmaf$ & $\sigma_{f-}$ & $\sigma_{f+}$ & $\Dsigmaf$   \\
     & [Hz] & [Hz] & [Hz] & [Hz] \\
    \hline
    \ac{gb}  & 0.076 & 0.078  & 0.074 & 0.0043   \\
    \ac{aips} & 0.042 & 0.0423  & 0.0433 & 0.0010   \\ 
    \ac{aus}  & 0.025 & 0.0267 & 0.024 & 0.0026 \\
    \ac{tas}  & 0.042 & 0.0458  & 0.0386 & 0.0072 \\
    \hline
  \end{tabular}
\end{table}

Regarding the asymmetry of the \ac{fpd}, $\Dsigmaf$, it is probably non-intuitive to see that \ac{aips} shows lower asymmetry compared to \ac{aus}, \ac{gb}, and \ac{tas} systems.  This is despite \ac{tas} and \ac{aus} systems having an \ac{agc} installed.  

\begin{figure}[htb]
  \subfigure[\ac{gb}]{\resizebox{0.495\linewidth}{!}{\includegraphics{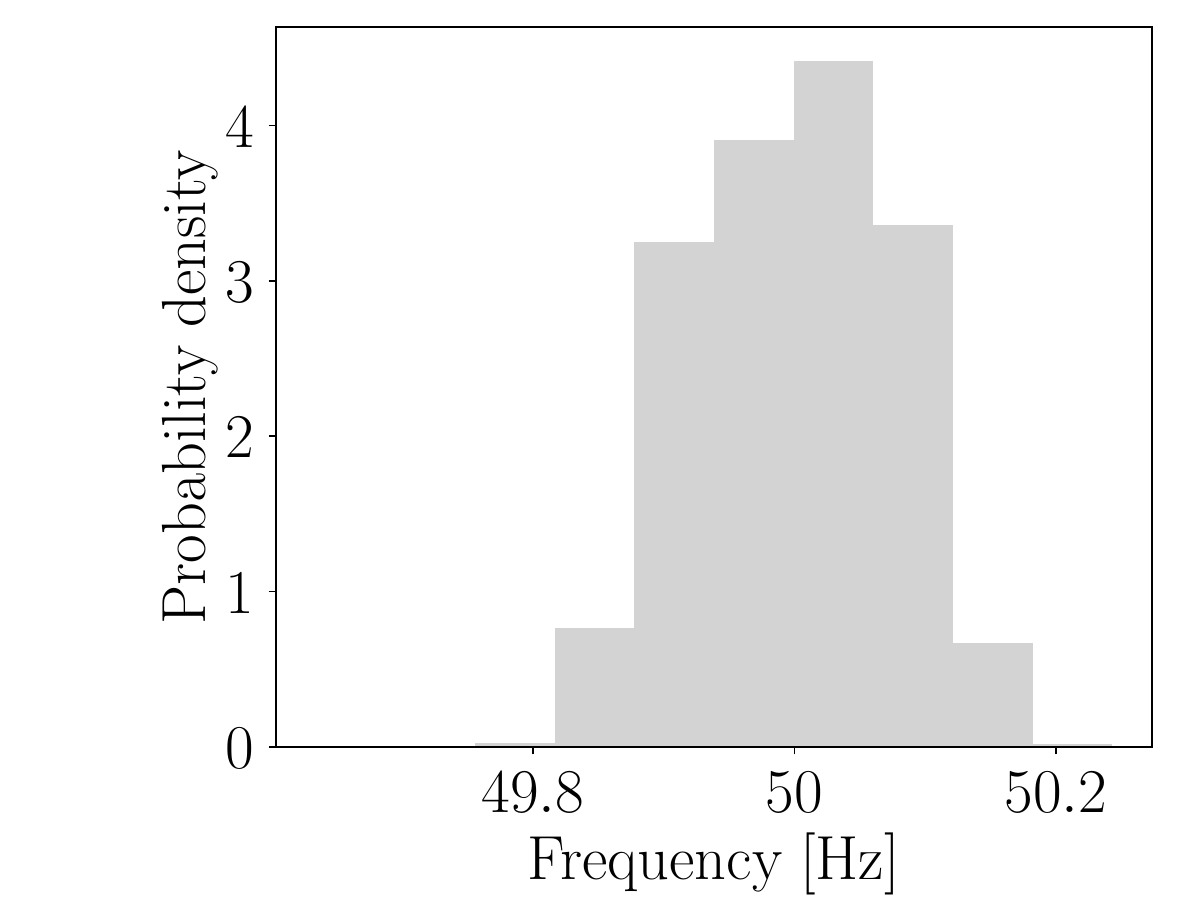}}}
  \subfigure[\ac{aips}]{\resizebox{0.495\linewidth}{!}{\includegraphics{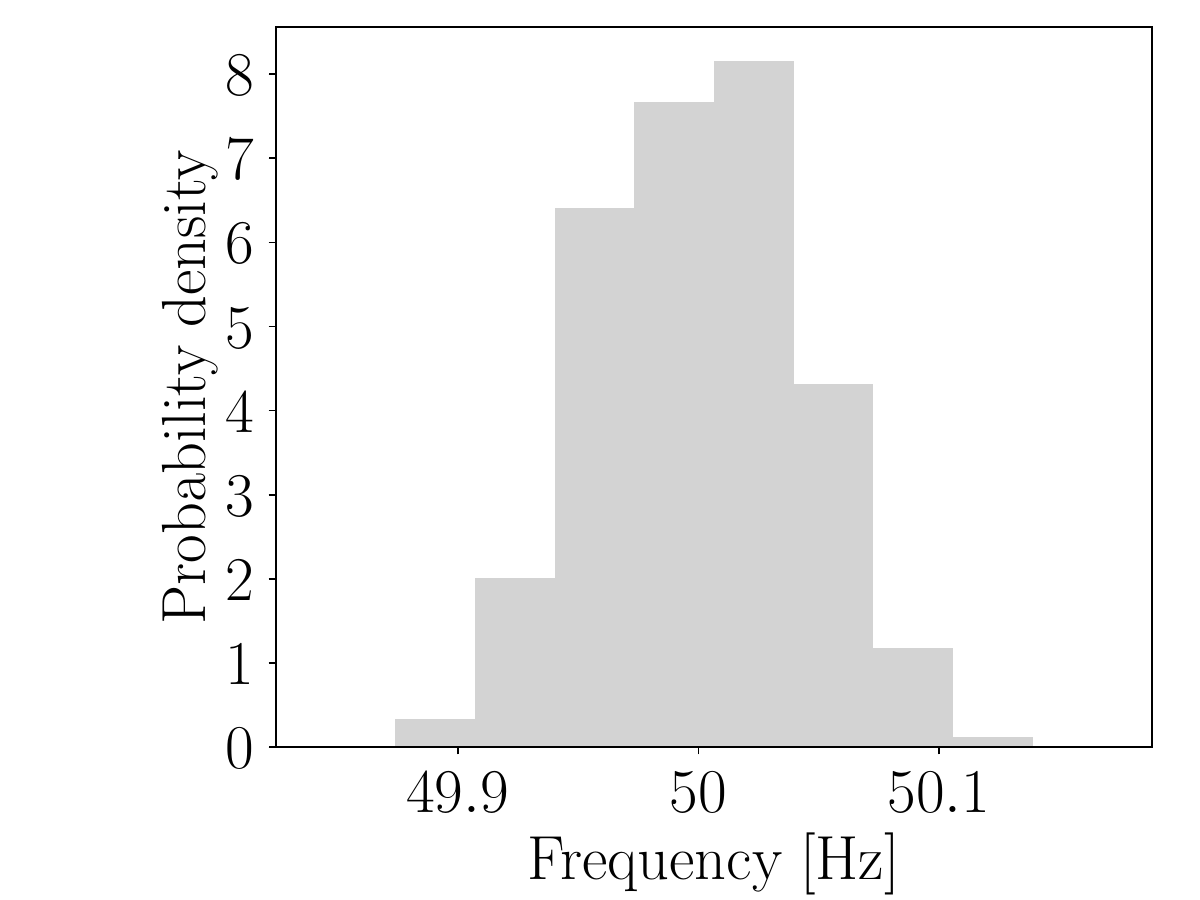}}}
  \subfigure[\ac{aus}]{\resizebox{0.495\linewidth}{!}{\includegraphics{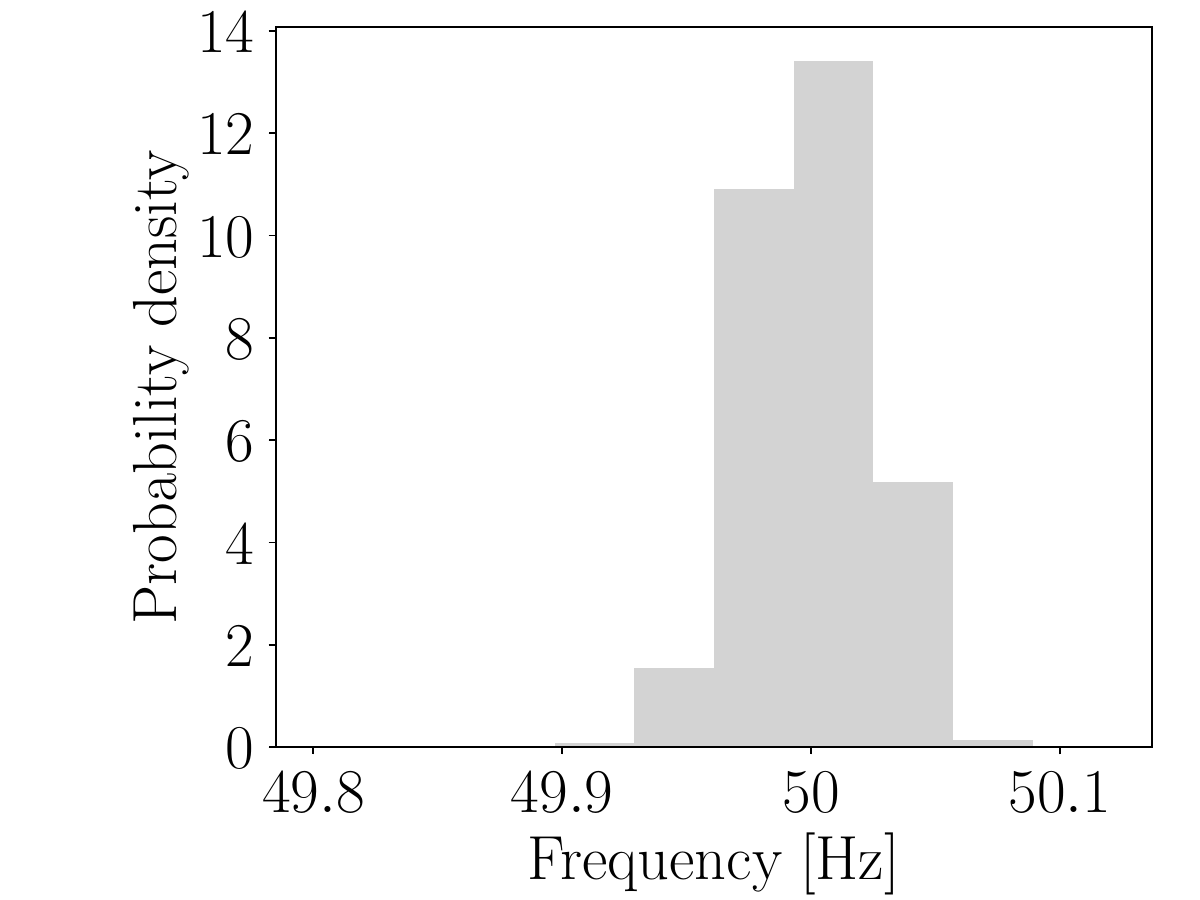}}}
  \subfigure[\ac{tas}]{\resizebox{0.495\linewidth}{!}{\includegraphics{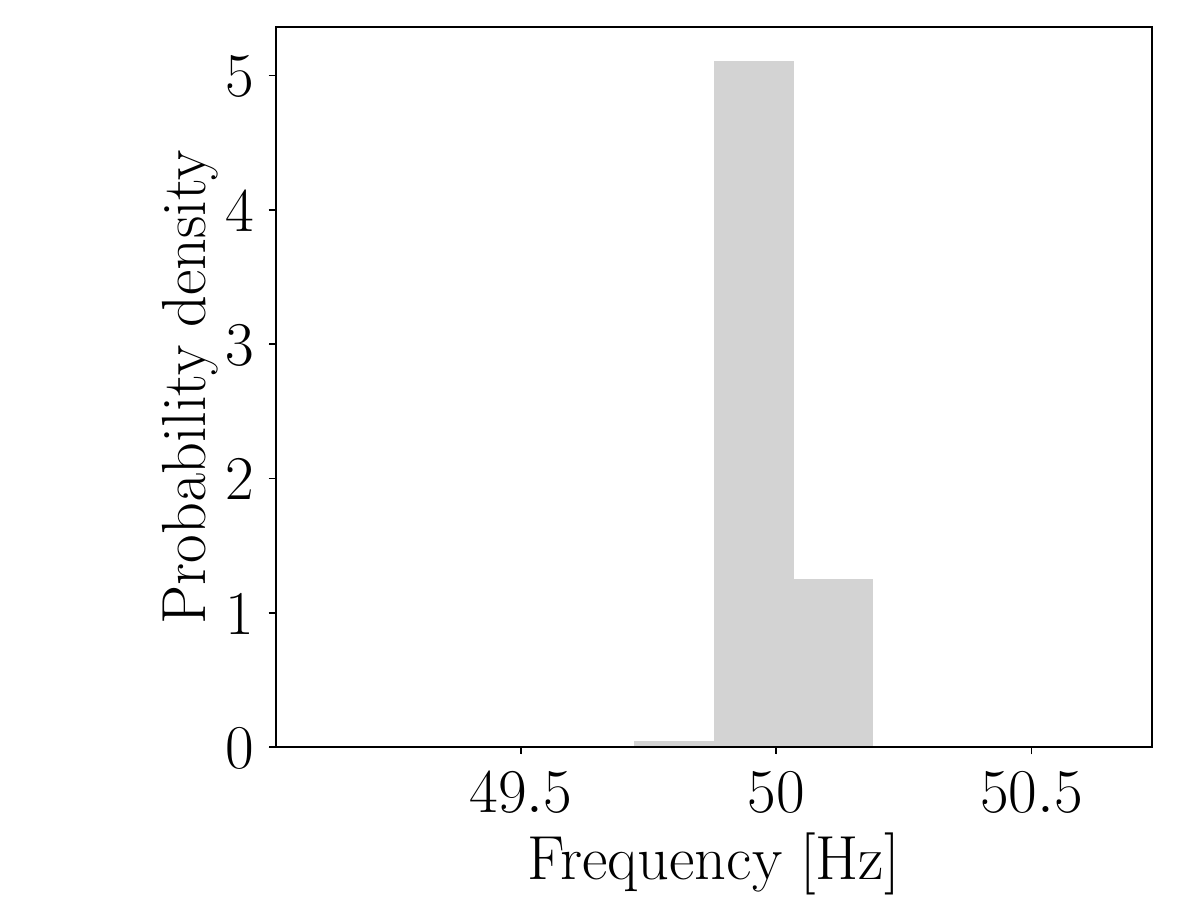}}} \\
  \caption[Comparison of FPD of four power systems in March 2024]{Comparison of \ac{fpd} of \ac{gb}, \ac{aips}, \ac{aus} and \ac{tas} power systems in March 2024.}
  \label{fig:05:pdf}
\end{figure}

Figure \ref{fig:05:pdf} shows the \acp{fpd} of the frequency of the four systems for March 2024.  Arguably, the source of asymmetry might come from \acp{ibr} providing mandatory \ac{pfc} with $\pm 15$ mHz dead-band in \ac{aus} and \ac{tas} similar to what is reported in \ac{aips} in \cite{kerci2024asymmetry}.  As mentioned in \cite{kerci2024asymmetry}, \ac{aemo} recognizes that the observed asymmetry in the \ac{nem}'s frequency characteristic could be due to the application of narrow dead-bands in some power plants \cite{aemopfc}.  

Note that in \ac{gb} 50 Hz is not the most common frequency.  For this reason, \ac{neso} procures frequency response assuming a pre-fault frequency different from 50 Hz \cite{NGESO}.  This approach is not common among other \acp{tso} that instead assume nominal frequency.  In any case, \ac{neso} maintains frequency within operational limits mainly because of the large size, that is, aggregated inertia, of its power system.  For example, it has been shown in the literature that the size of the grid serves as a controlling factor to make grid dynamics more robust \cite{schafer2018non, tchuisseu2023secondary}.  In other words, ``size'' can be seen as part of the frequency control and, in this case, might ``hide'' the strength, or lack thereof, of control.  This aspect is further discussed in the next section.

\subsection{Abnormal System Conditions}
\label{sec:05:may2}

We now discuss the frequency control strength for abnormal operating conditions.  Tables \ref{tab:05:cont1} and \ref{tab:05:cont2} compare: (i) information on recent relevant trips of the four systems; (ii) operating conditions, namely, inertia, total demand at the time and power imbalance due to the outage; and (iii) various frequency strength metrics as well as the five \ac{rocof}-based metrics described above. 
 
\paragraph{\ac{gb} and \ac{aips} trips on May 14, 2024}

On May 14, 2024, at around 1 am, the two \ac{hvdc} \acp{ic} that connect \ac{aips} with \ac{gb} tripped one 20 s after the other with a total import (\ac{aips}) / export (\ac{gb}) capacity lost of 912 MW (530 MW and 382 MW).  This allows an excellent frequency control strength comparison, as it represents the same contingency for the two systems.  Tables \ref{tab:05:cont1} and \ref{tab:05:cont2} show system conditions and all relevant results for the \ac{aips} 2024 and \ac{gb} 2024 cases, while Figure \ref{fig:05:cont} shows the relevant frequency transients following the \acp{ic} trips.  In particular, note that the inertia level in \ac{gb} was just above the limit (121 GWs) while in \ac{aips} was quite above the limit (32.3 GWs) as wind levels were particularly low (534 MW) which means many conventional units online (2153 MW) to meet demand (3699 MW).  Also, note that only the MW trip of the first \ac{ic} (${\rm \Delta} p = 530$ MW) is shown in the table as that is important for the different \acp{rocof} calculations.  The second \ac{ic}, in fact, trips 20 s later and less active power is lost.  

Tables \ref{tab:05:cont1} and \ref{tab:05:cont2} show that frequency zenith/nadir for the \ac{gb}/\ac{aips} systems reached 50.154 Hz and 49.593 Hz, respectively.  The difference in the maximum frequency deviations is expected as the total active power trip represented around 25\% of total demand for \ac{aips} system while only around 4\% for \ac{gb}.  Also, nadir happens quickly in \ac{aips} (2 s for the first trip and 26 s for the second trip) than zenith in \ac{gb} (4 s and 27 s, respectively) due to much lower inertia.  Frequency took around 969 s to recover within $\pm 100$ mHz for \ac{aips} system while for \ac{gb} only 393 s.  Results also show that the maximum calculated \acp{rocof} ($\rocofmax$) based on 1 s (4 s) resolutions are 0.066 Hz/s (0.034 Hz/s) and 0.16 Hz/s (0.06) Hz/s for \ac{gb} and \ac{aips} systems, respectively.  These absolute values are expected considering the different inertia levels for both systems namely 121 GWs and 32.3 GWs for \ac{gb} and \ac{aips}, respectively.  What is interesting, though, is the fact that the $\rocofmax$ for \ac{gb} (0.066 Hz/s) almost matches the initial $\rocofH = 0.063$ Hz/s.  This suggest that the inertial level of \ac{gb} appears to be a good indication of the $\rocofmax$ experienced even when using 1 s resolution data.  

\begin{table*}[t!]
  \centering
  \caption[Frequency strength for contingency events]{Frequency strength for contingency events in \ac{gb}, \ac{aips}, \ac{aus} and \ac{tas}.}
  \label{tab:05:cont1}
  \begin{tabular}{llcccccc}
    \hline
    Power system & & \ac{gb} & \ac{aips} & \ac{gb} & \ac{aips} & \ac{aus} & \ac{tas} \\
    Year & & 2024 & 2024 & 2019 & 2022 & 2024 & 2024 \\
    \hline
    Inertia & [GWs] & 121 & 32.3 & 201 & 34.7 & 90 & 6 \\
    $p_{\rm conv}$ & [MW] & 9,649 & 2,153 & 15,980 & 3,077 & -- & --\\
    $p_{\rm total}$ & [MW] & 21,773 & 3,699 &  28,029 & 4,847 & 23,122 & 1,111  \\
    ${\rm \Delta} p_{\rm imbalance}$ & [MW] & 530 & 530 & 1,000 & 530 & 660 & 114\\
    \hline
    Nadir/zenith & [Hz] & 50.154 & 49.593 & 49.62 & 49.674 & 49.8 & 49.3 \\
    Time to nadir/zenith & [s] & 4/27 & 2/26 & 6 & 3  & 8 & 8\\
    Time to recover within $\pm100$ mHz & [s] & 393 & 969 & 255 & 780 & 64 & 48 \\
    \hline
  \end{tabular}
\end{table*}

\begin{table*}[t!]
  \centering
  \caption[RoCoF-based metrics for contingency events]{RoCoF-based metrics for contingency events in \ac{gb}, \ac{aips}, \ac{aus} and \ac{tas}.}
  \label{tab:05:cont2}
  \begin{tabular}{llccccccccccc}
    \hline
    Power system & & & \ac{gb} & & & \ac{aips} & & & \ac{gb} \\
    Year & & & 2024 & & & 2024 & & & 2019 \\
    \hline
    $i$ & [s]& 1 & 4 & 0.02 & 1 & 4 & 0.02 & 1 & 4 & 0.02 \\
    \hline
    $\rocofH$  & [Hz/s]  & & 0.063 &&& 0.41 &&& 0.124 \\
    $\rocofavg$  & [Hz/s]  & 0.0021 & 0.0012& -- & 0.0012& 0.0010& -- & 0.0026& 0.0022& -- \\
    $\rocofmax$   & [Hz/s] & 0.066 &0.034& -- & 0.16& 0.06& 0.27 & 0.118& 0.085& -- \\
    $\rocofra$ & [Hz/s] & 2.71& 1.39& -- & 1.11& 0.42& 1.88 & 3.30& 2.38& -- \\
    $\rocofrb$ & [Hz/s] & 1.20&0.62& -- & 0.65&0.24& 1.09 & 1.88& 1.36& -- \\
    \hline
    \\
    \hline
    Power System & & & \ac{aips} & & & \ac{aus} & & & \ac{tas} \\
    Year & & & 2022 & & & 2024 & & & 2024 \\
    \hline
    $i$ & [s]& 1 & 4 & 0.02 & 1 & 4 & 0.02 & 1 & 4 & 0.02 \\
    \hline
    $\rocofH$  & [Hz/s]  & & 0.38 &&& 0.18 &&& 0.47  \\
    $\rocofavg$ & [Hz/s] & 0.0018 & 0.0012 & -- & -- & 0.0024 & -- & -- & 0.0072 & -- \\
    $\rocofmax$ & [Hz/s] & 0.242 & 0.086 & -- & -- & 0.023 & 0.14 & -- & 0.23 & 0.46 \\
    $\rocofra$ & [Hz/s] & 2.21 & 0.78 & -- & -- & 0.80 & 4.90 & -- & 2.24 & 4.48 \\
    $\rocofrb$ & [Hz/s] & 1.40 & 0.50 & -- & -- & -- & -- &  -- & -- & -- \\
    \hline
  \end{tabular}
\end{table*}

\begin{table}[t!]
  \centering
  \caption[FFR volumes in four power systems]{\ac{ffr} volumes in \ac{gb}, \ac{aips} and \ac{aus} systems.}
  \label{tab:05:ffrvol}
  \begin{tabular}{ccccc}
    \hline
    Direction & Units & \ac{gb} (\ac{dc}) & \ac{aips} (\ac{ffr}) & \ac{aus}/\ac{tas} (\ac{ffr})    \\
    \hline
    Upward & [MW] & $\sim$ 1,300 & $\sim$ 1,800 & 250  \\
    Downward & [MW] & $\sim$ 1,300 & $<$ 100 & 125   \\
    \hline
  \end{tabular}
\end{table}

\begin{figure}[thb!]
  \begin{center}
    \resizebox{1.0\linewidth}{!}{\includegraphics{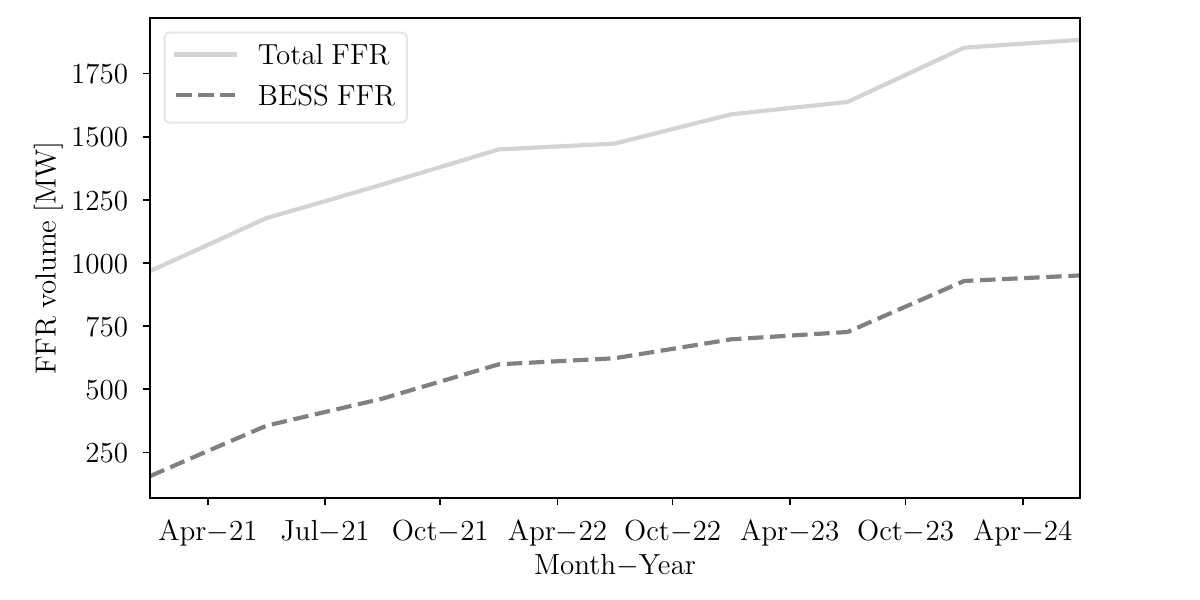}}
    \caption[Evolution of FFR contracted volumes in the AIPS]{Evolution of \ac{ffr} contracted volumes in the \ac{aips}.}
    \label{fig05:ffr_volumes}
  \end{center}
\end{figure}

For the \ac{aips} system, on the other hand, $\rocofmax = 0.16$ Hz/s (while the one calculated from 20 ms resolution data and over a rolling 500 ms period is higher namely 0.27 Hz/s, as expected).  The initial \ac{rocof} calculated with \eqref{eq:05:rocof2} equals $\rocofH = 0.41$ Hz/s (still higher than 0.27 Hz/s from 20 ms resolution data).  The fact that $\rocofH > \rocofmax$ means that there is more frequency support in the inertial time frame in the \ac{aips} than in \ac{gb}.  This can be explained by significant volumes of \ac{ffr} being provided in less than a second in the \ac{aips} system and thus helping with addressing \ac{rocof} as well (instead \ac{gb} has a full \ac{ffr} delivery requirement of 1 s).  For example, Table \ref{tab:05:ffrvol} and Figure \ref{fig05:ffr_volumes} show that the \ac{aips} system has around 1,800 MW of upward \ac{ffr} available (tariff-based procurement) to deal with \ac{uf} events.  This volume is significantly higher compared to its size and the rest of the power systems and thus comes with a significant cost.  In this context, \ac{neso} has developed a clear and transparent methodology to determine
the right balance between the two competing objectives of reliability and cost, focusing on the
risks, impacts and controls for managing the frequency \cite{NESOrocof}.  

\begin{table}[htb]
  \centering
  \caption[Assessment of minimum inertia requirements in GB]{Assessment of minimum inertia requirements in \ac{gb}.} 
  \label{tab:05:gbinertia}
  \begin{tabular}{lcccc}
    \hline
    Inertia floor [GWs] & 140 & 120 & 110 & 102  \\
    Cost [£m] & 524 & 266 & 198 & 170 \\
    \hline
  \end{tabular}
\end{table}

For instance, the 2025 \ac{frcr} from \ac{neso} recommends reducing the inertia floor from 120 GWs to 102 GWs due to significant cost savings (see Table \ref{tab:05:gbinertia} \cite{NESOinertiafloor}).  To improve system risk, \ac{neso} recommends procuring 200 MW additional \ac{dc}-Low (or upward \ac{ffr}) as the most cost-effective solution.

In fact, because of high reserve costs, the \ac{aips} system is introducing more competitive arrangements to procure reserves (auction-based) in the future including \ac{ffr} \cite{fass}.  Another factor could be that there might be slightly more inertia available in the \ac{aips} system than the 32.3 GWs figure (assumed coming only from conventional generators and neglecting inertia from demand, for example), while for \ac{gb} the 121 GWs figure might be representing better the actual inertia in the system.

Based on the discussion above and the considered metrics, it appears that the \ac{gb} power system is stronger than the \ac{aips} system in absolute terms.  However, if we calculate the relative frequency strength of the two power systems using equations \eqref{eq:05:deltaw} and \eqref{eq:05:deltaw1} the situation changes.  Specifically, $\rocofra$ values for \ac{gb} and \ac{aips} systems are 2.71 Hz/s (1.39 Hz/s) and 1.11 Hz/s (0.42 Hz/s), respectively, and thus lower for \ac{aips}.  Note that the values before (within) brackets are calculated using $\rocofmax$ from 1 s (4 s) data resolution.  Similarly, the $\rocofrb$ values are lower for \ac{aips} than \ac{gb} namely 0.65 Hz/s (0.24 Hz/s) and 1.20 Hz/s (0.62 Hz/s).  

\begin{figure}[htb]
  \subfigure[\ac{tas}: March 6, 2024]{\resizebox{0.495\linewidth}{!}{\includegraphics{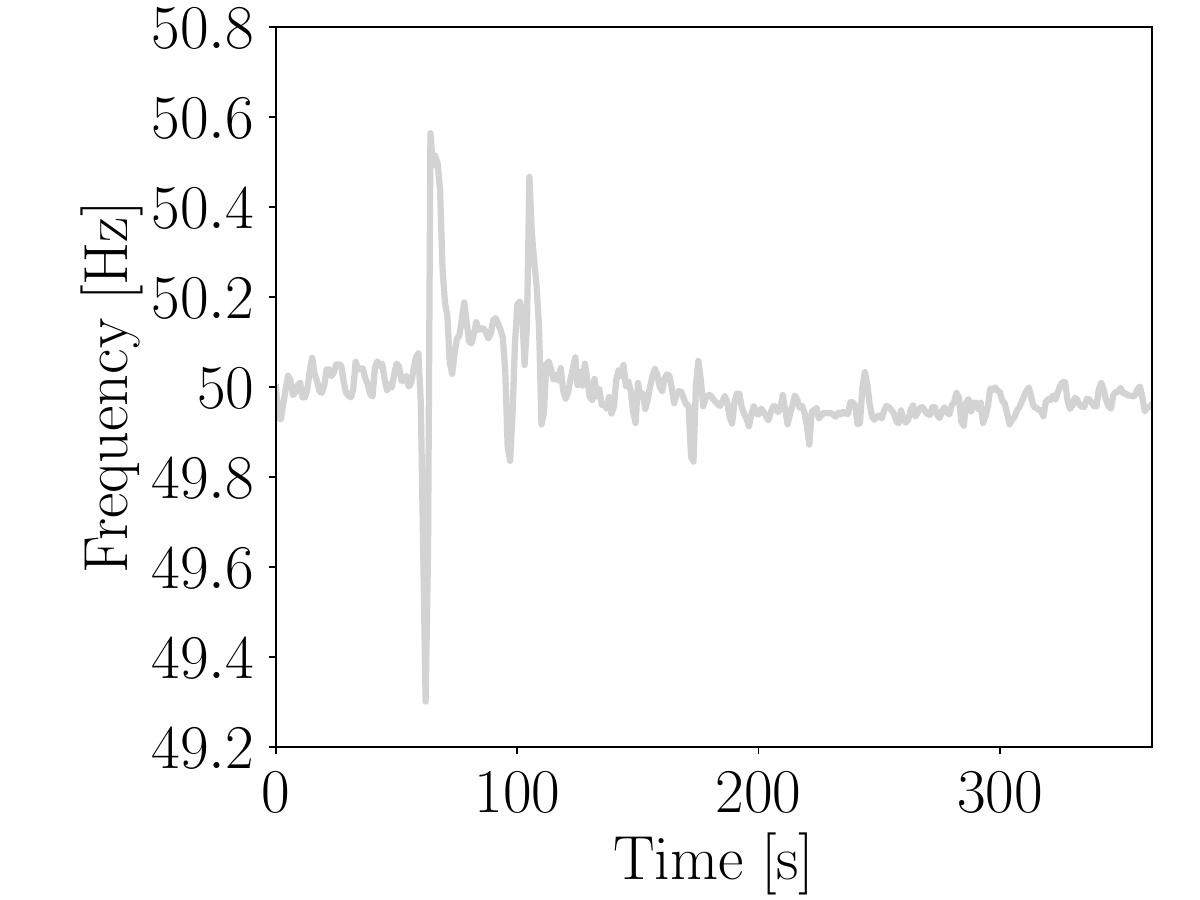}}}
  \subfigure[\ac{aus}: March 21, 2024]{\resizebox{0.495\linewidth}{!}{\includegraphics{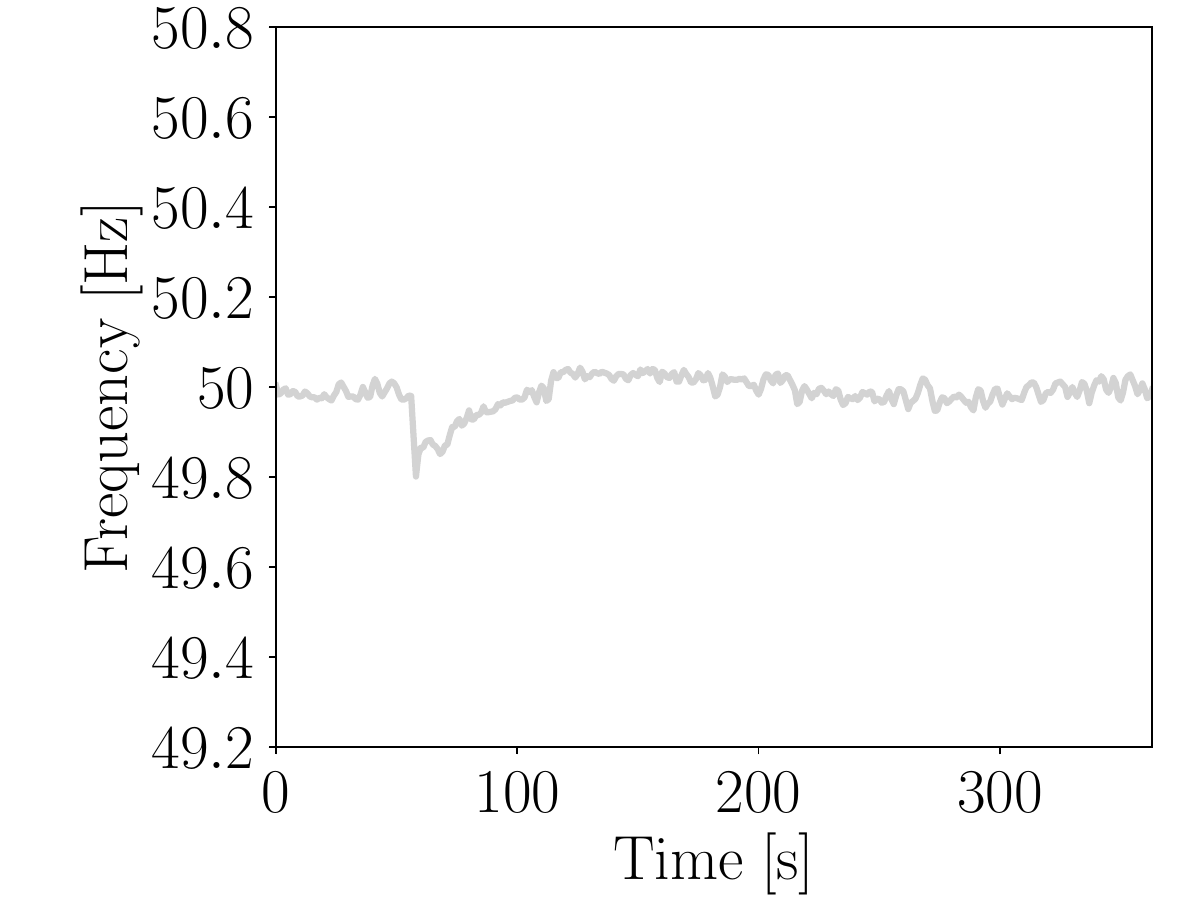}}}
  \subfigure[\ac{aips}: May 14, 2024]{\resizebox{0.495\linewidth}{!}{\includegraphics{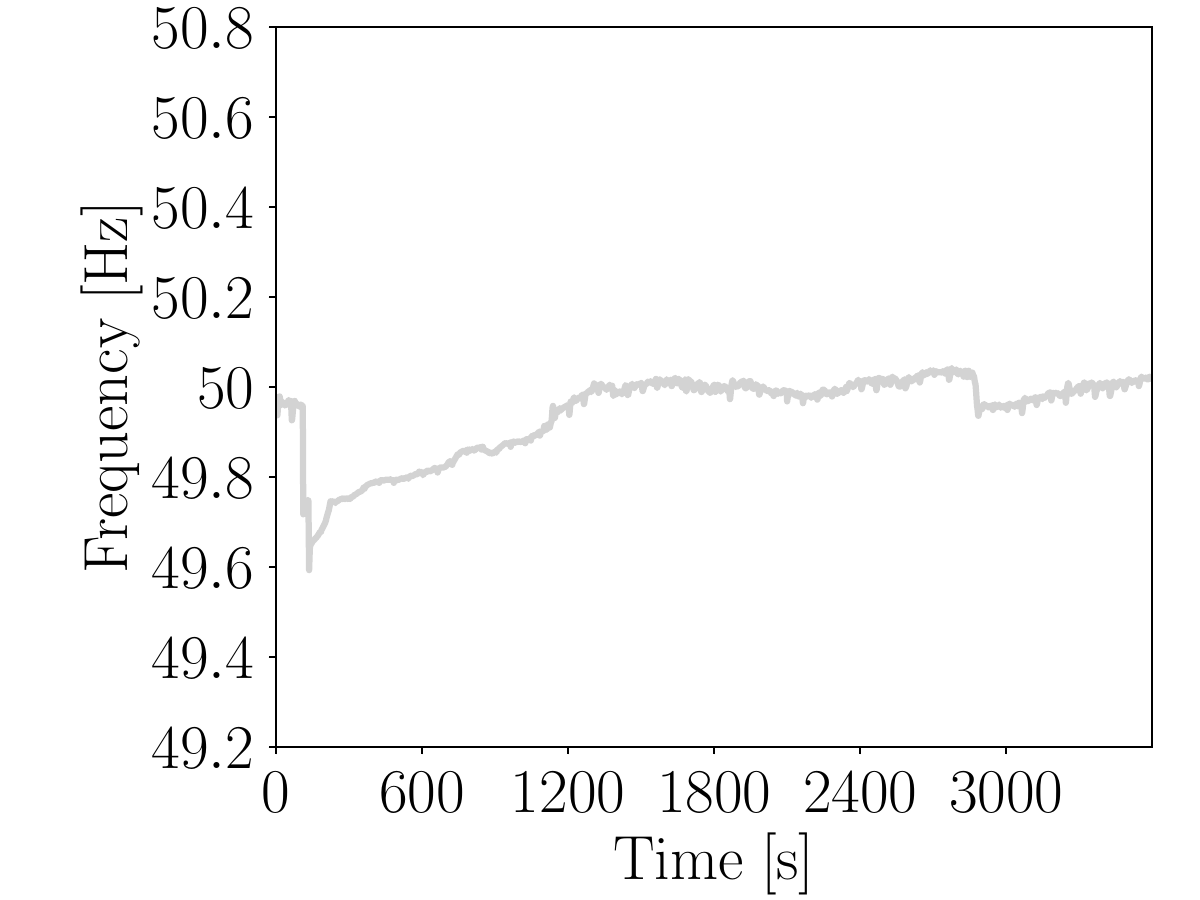}}}
  \subfigure[\ac{gb}: May 14, 2024]{\resizebox{0.495\linewidth}{!}{\includegraphics{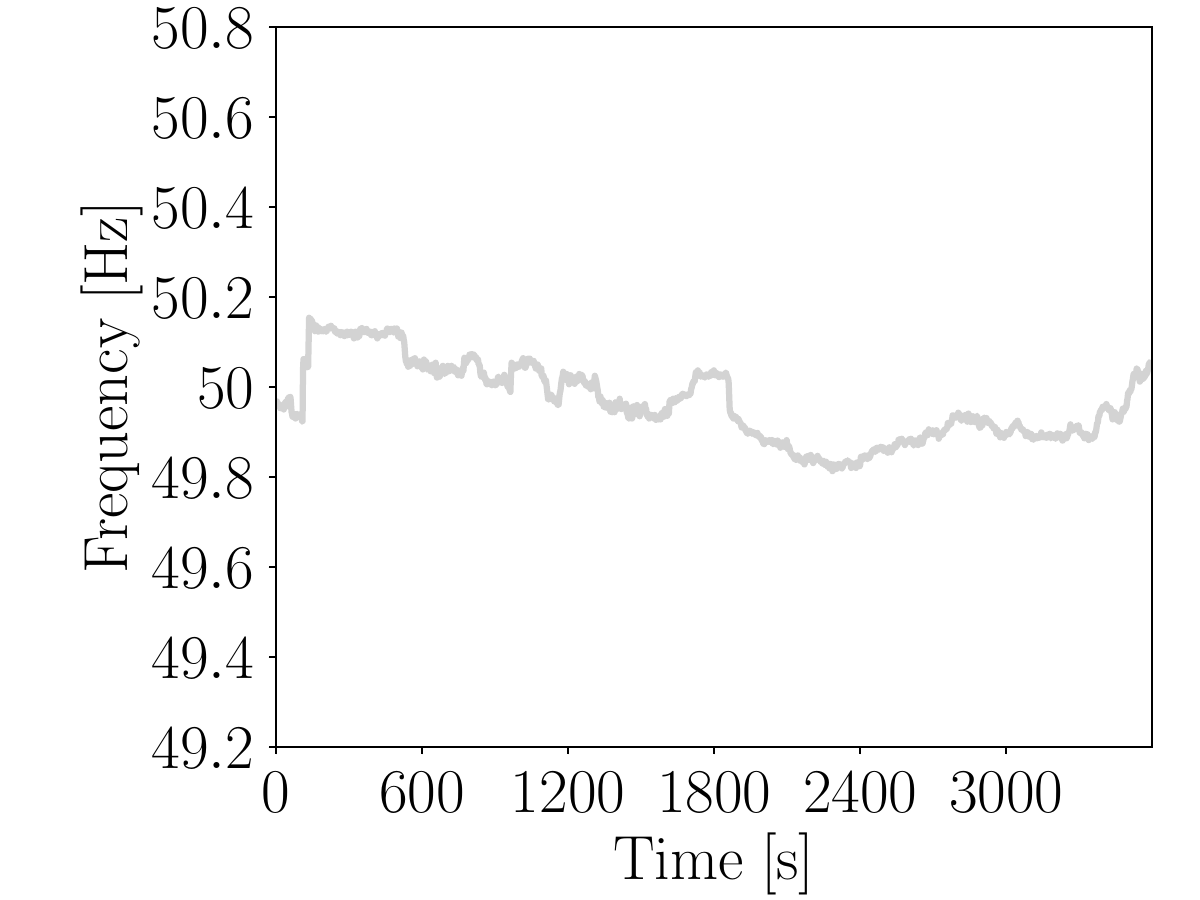}}} \\
  \caption[Comparison of frequency traces of four power systems in 2024]{Comparison of frequency traces of the \ac{gb}, \ac{aips}, \ac{aus}, and \ac{tas} power systems for relevant contingencies in 2024.}
  \label{fig:05:cont}
\end{figure}

\paragraph{\ac{gb} trip on July 1, 2019 and \ac{aips} trip on August 9, 2022}

To further validate the above results and conclusions, we select two more relevant contingencies for \ac{gb} ($\rm GB_{2019}$) and \ac{aips} ($\rm AIPS_{2022}$) namely \ac{ic} trip (1,000 MW import) on July 1, 2019 at 08:27 in \ac{gb} and \ac{ic} trip (530 MW import) on August 9, 2022 in \ac{aips} \cite{NGESOreport}.  The respective system operating conditions and results are given in Table \ref{tab:05:cont1}.  It can be seen that the results are very similar to the previous comparison.  Specifically, the nadir happens quicker in \ac{aips} than in \ac{gb} and similarly the frequency restoration in \ac{aips} takes much more time.  The \ac{gb} system performs better than the \ac{aips} system in terms of frequency recovery following contingencies.  

Next, notice again that $\rocofmax$ and $\rocofH$ for \ac{gb} are very similar (0.118 Hz/s and 0.124 Hz/s, respectively) while that is not the case for \ac{aips} (0.242 Hz/s and 0.38 Hz/s, respectively).  As mentioned above, two contributing factors here are the bigger support from \ac{ffr} in the \ac{aips} system (response time down to 150 ms) and potentially more inertia available than the calculated inertia from conventional units in \ac{aips} (34.7 GWs inertia figure might be higher in reality) compared to \ac{gb} (201 GWs seems to be close to actual inertia).  It is also fascinating to see that the difference between $\rocofmax$ and $\rocofH$ has increased for \ac{aips} over the last years (0.138 Hz/s in 2019 and 0.25 Hz/s in 2024).  Basically, this means that the frequency support in the inertial time frame has increased steadily (see Figure \ref{fig05:ffr_volumes}).  This, in turn, has led to significant frequency stability improvement in recent years in the \ac{aips} system \cite{10253411}.  

With regard to the relative \ac{rocof} calculations, again, \ac{aips} shows lower values than \ac{gb}.  Specifically, $\rocofra$ values for \ac{gb} and \ac{aips} systems are 3.30 Hz/s (2.38 Hz/s) and 2.21 Hz/s (0.78 Hz/s), respectively.  $\rocofrb$ values are also lower for \ac{aips} (1.40 Hz/s (0.50 Hz/s)) than \ac{gb} (1.88 Hz/s (1.36 Hz/s)).  This lead to confirm that the \ac{aips} is stronger than the \ac{gb} system relative to its size.  Considering both jurisdictions are moving towards lower levels of inertia (Table \ref{tab:05:gbinertia}) and competitive procurement of reserves, it would be interesting to perform a similar comparison in a few year's time to see whether the above conclusion will still hold. 

\paragraph{\ac{aus} trip on March 21, 2024 and \ac{tas} trip on March 6, 2024}

We conclude the comparison for abnormal operating conditions by selecting two relevant trips for \ac{aus} and \ac{tas} systems namely March 21, 2024 (trip of Bayswater Power Station Unit 1 at 660 MW at 18:00) and March 6, 2024 (trip of Comalco at 114 MW at 8:07), respectively \cite{aemo3}.  It is important to note that the frequency measurements for these two events are provided in 4 s resolution and the inertia values are guessed based on time series graphs provided in \cite{aemo3}.  While these minor limitations in data might lead to small discrepancies, we note that it does not affect the main conclusions drawn. 

The system operating conditions and results are provided in Table \ref{tab:05:cont1}, while Figure \ref{fig:05:cont} illustrates the relevant frequency transients.  Due to the size of the \ac{tas} system (around 6 GWs compared to 90 GWs of \ac{aus}) and the size of the contingency at the time of incident (114 MW representing approximately 10\% of demand), frequency nadir reached the lowest value (49.3 Hz).  Regarding the time to nadir, both the \ac{aus} and \ac{tas} systems show a similar value of 8 s.  Note, however, that since only 4 s data is available these times may be slightly lower and thus similar to \ac{aus} and \ac{aips} systems.  On the other hand, in contrast to \ac{gb} and \ac{aips}, the \ac{aus} and \ac{tas} power systems show a much better frequency recovery, namely, tens of seconds vs hundreds of seconds for \ac{gb} and \ac{aips}.  This could be explained by the fact that both \ac{aus} and \ac{tas} systems employ an \ac{agc} while that is not the case for \ac{gb} and \ac{aips} systems (see Table \ref{tab05:param1}).  

Regarding different \acp{rocof} calculations, one can see that the average $\rocofavg$ ($i=4$) shows a higher value for \ac{aus} (0.0024 Hz/s) and \ac{tas} (0.0072 Hz/s) systems as compared to the \ac{gb} (0.0012 Hz/s and 0.0022 Hz/s) and \ac{aips} (0.0010 Hz/s and 0.0012) systems.  This makes sense considering that both the \ac{aus} and \ac{tas} systems recover their frequency quite quickly to $\pm 100$ mHz (see Figure \ref{fig:05:cont}).  With regard to the maximum $\rocofmax$, since there is no 1 s resolution data, we utilize the value provided in \cite{aemo3} by \ac{aemo} (0.14 Hz/s and 0.46 Hz/s for \ac{aus} and \ac{tas}, respectively).  \ac{aemo} calculates these from high-resolution data (20 ms) and 500 ms rolling window and filtering short-term transients \cite{aemo3}.  But if we are to calculate $\rocofmax$ from 4 s resolution data then these values are 0.023 Hz/s and 0.23 Hz/s for \ac{aus} and \ac{tas}, respectively.  This means that the \ac{tas} system experiences the worst $\rocofmax$ (0.23 Hz/s) compared to \ac{gb} (0.034 Hz/s and 0.085 Hz/s), \ac{aips} (0.06 Hz/s and 0.086 Hz/s), and \ac{aus} (0.023 Hz/s) systems.  Similar to the nadir explanation above, this is to be expected considering the size of the \ac{tas} system and contingency.

Looking at the $\rocofH$ calculations, we can see that they are very similar to those calculated using 20 ms data and 500 ms rolling window namely 0.14 Hz/s and 0.18 Hz/s for \ac{aus}, and 0.46 Hz/s and 0.47 Hz/s for \ac{tas} system.  Similar to the \ac{gb} cases ($\rm GB_{2024}$ and  $\rm GB_{2019}$), the similarity in the values of $\rocofH$ and $\rocofmax$ indicate that the inertial levels of the \ac{aus} and \ac{tas} systems are a good indication of the actual experienced \ac{rocof} and that there is little \ac{ffr} being provided in the inertial time frame.  Note that even if the inertia values of the \ac{aus} and \ac{tas} systems might be slightly different than the current values (90 GWs for \ac{aus} and 6 GWs for \ac{tas}), the value of the $\rocofH$ will not change significantly and thus will not affect the main conclusions.  

The $\rocofra$ results mean that the \ac{aus} (0.80 Hz/s) shows higher relative frequency control strength compared to \ac{gb} (1.39 Hz/s and 2.38 Hz/s) and \ac{tas} (2.24 Hz/s) systems and lower strength compared to the \ac{aips} (0.42 Hz/s and 0.78 Hz/s) system.  Therefore, it can be concluded that based on the relative \acp{rocof} comparisons namely $\rocofra$ and $\rocofrb$, the \ac{aips} system shows a higher relative frequency control strength than \ac{gb}, \ac{aus} and \ac{tas} systems, in fact, relative $\rocofra$ and $\rocofrb$ are lower for \ac{aips}.  Note that if one considers other frequency strength metrics such time to recover within $\pm 100$ mHz, then the above conclusion may completely change.  However, we think that since \ac{rocof}-based results, in particular, $\rocofra$ and $\rocofrb$, are more critical/important than the recovery period in terms of frequency stability, it makes sense to reach the conclusion based on those results.

\section{Conclusions}
\label{sec:05:conclu}

The common understanding is that the bigger the capacity of a power system, the bigger its robustness with respect to events and contingencies.  Data discussed in this paper show that this is not always the case in the context of frequency control.  Specifically, the key findings of this paper are summarized below.

\subsection{Normal operating conditions}

Our analysis indicates that despite being the second-biggest power system, \ac{aus} performs better in frequency regulation during normal system conditions.  This is a counterintuitive result from a system size point of view \cite{schafer2018non} and, in particular, because \ac{neso} has recently introduced two new dynamic frequency regulation products (but \ac{dc} also helps slightly in normal conditions by having $\pm 15$ mHz dead-band \cite{NGESO}).  A possible mitigation for \ac{gb} is to consider a similar path to \ac{aus} regarding mandatory \ac{pfc} provision with a narrow dead-band.  If this is the case, it is suggested to focus on a potential increase in the asymmetry of frequency distribution like in \ac{aus}.  Alternative solutions include implementing an \ac{agc} and revising frequency regulation products.  Despite the \ac{aus} system showing, overall, the best frequency regulation performance, it appears that frequency asymmetry ($\Dsigmaf$) is higher than in \ac{aips}.  It is suggested to study more in detail $\Dsigmaf$ and its sources in the \ac{aus} system.

\subsection{Abnormal operating conditions}

The frequency control strength comparison is more complex during contingency events due to the different stages of frequency control such as \ac{ffr}/\ac{pfc} and \ac{agc}.  The \ac{aips} power system appears the ``strongest'' to arrest frequency and \ac{rocof} relative to its size.  This is mainly due to the significant procurement of \ac{ffr} volumes and thus comes to a higher cost compared to the rest of the power systems.  The \ac{aus} and \ac{tas} show better frequency recovery compared to the \ac{aips} and \ac{gb} power systems.  A possible solution to this problem could be that \ac{gb} and \ac{aips} consider implementation of automatic secondary frequnecy control such as \ac{agc} similar to \ac{aus} and \ac{tas}.


\vfill

\end{document}